\documentclass[aps,twocolumn,prb,amsmath,amssymb,showpacs,floatfix,superscriptaddress,
]{revtex4-2}

\usepackage{graphicx}
\usepackage{changes}
\usepackage{times}
\usepackage[varg]{txfonts}
\usepackage{bm}
\usepackage{amsmath}
\usepackage{sansmath}
\usepackage{xspace}
\usepackage{xr}
\usepackage{natbib}
\usepackage{xcolor}
\usepackage{subfig}
\usepackage{subfig}
\usepackage{appendix}

\usepackage{tabularx}
\usepackage{setspace}
\usepackage{booktabs}
\usepackage{braket}
\usepackage{csquotes}
\usepackage{wrapfig}

\usepackage[colorlinks,bookmarks =false,citecolor=blue,linkcolor=red,urlcolor=blue]{hyperref}
\usepackage{cleveref}

\usepackage{todonotes} 
\usepackage{array}

\newcommand{\rim}[1]

\newcommand{\xUZH}{Department of Physics, University of Z{\"u}rich, 8057 Z{\"u}rich, Switzerland}

\newcommand{\xSUStech}{Department of Physics, Southern University of Science and Technology (SUSTech), Shenzhen 518055, China}

\begin{document}

\title{Non-linear transport in multifold semimetals}

\author{Andrea Kouta Dagnino}
\affiliation{\xUZH}

\author{Xiaoxiong Liu}
\affiliation{\xSUStech}

\author{Titus Neupert}
\affiliation{\xUZH}

\begin{abstract}
Transport measurements are a powerful way to probe the electronic structure of quantum materials, but the information they contain is often convoluted. Yet, in particular for simple low-energy fermiologies, and by combining linear and non-linear responses, definite conclusions can be drawn -- such as, for instance, in the case of the circular photogalvanic effect in Weyl semimetals.
Here, we derive the complete DC transport response functions up to third order in the applied electric field within Boltzmann theory that encode combined information about quantum geometry and band dispersion. We discuss the responses for multifold fermions at high-symmetry momenta in time-reversal symmetric crystals as well as their reduction by symmetry constraints. We exemplify in detail the cases of space group $P2_13$ and $I4_132$, which realize different multifold fermions, and show under which conditions these low-energy excitations can be differentially addressed through their bulk non-linear responses, enabling non-linear valley-tronics. We show that in the latter space group, a purely transverse third order valley current can be produced, leading to an accumulation of electrons of different valleys on opposite surfaces of the material.
\end{abstract}

\maketitle

\section{Introduction}\label{sec:introduction}
The physics of electrons in crystals is most directly distinguished from free electrons through the change in their dispersion relation imprinted by the lattice potential, with features such as van Hove singularities, relativistic fermions~\cite{bradlyn2016beyond}, or nested Fermi surfaces. In recent years a second lattice effect, namely the momentum dependence of Bloch states -- subsumed under the term quantum geometry -- came into focus~\cite{RevModPhys.82.1959, bzdusek2025quantumgeometrynonlinearoptics}. It can profoundly affect a material's response functions as well as possible interacting instabilities and can relate to topological properties. 

Insights about quantum geometry can often be drawn from a real-space perspective, where it affects the localization and shape of Wannier orbitals~\cite{PhysRevB.83.035108}. Overlaps of such Wannier orbitals, in turn, determine the band-projected (i.e., effective low energy) interaction matrix elements with implications for superconducting instabilities~\cite{PhysRevLett.128.087002, PhysRevLett.131.240001}, electron-phonon coupling~\cite{Bernevig-el-ph} or topological liquids~\cite{PhysRevB.86.241104,PhysRevLett.133.156504, CRPHYS_2013__14_9-10_816_0}. The real-space perspective can also be suitable for understanding quantum geometry contributions to electronic responses, in particular the linear and non-linear optical conductivity. Examples are: (i)~the topological Hall conductivity of band insulators, which is accompanied by an  obstruction against exponential localization of Bloch orbitals~\cite{DJThouless_1984}. (ii)~The shift-current in noncentrosymmetric materials, which is a second order response following from a dynamical displacement of Wannier centers at first order~\cite{Cook2017, PhysRevB.110.075159}. (iii) The related circular photogalvanic effect (CPGE) with a universally quantized response coefficient in three-dimensional (3D) Weyl semimetals~\cite{Juan2017}. Related optical probes of topological semimetals include the circular dichroism in higher-pseudospin semimetals and strong-field non-linear optical signatures of multi-Weyl nodes \cite{SekhMandal2022,BhartiDixit2023}. Furthermore, it has been recently shown that under suitable conditions on the band structure, the frequency integrated shift current response is determined by a \emph{multiband} topological invariant~\cite{PhysRevLett.133.186601} -- the most direct physical implication of such a topological descriptor known to date.

Despite the wealth of relations between quantum geometry and charge conductivities, experiments that unambiguously isolate the quantum geometric origin of response functions are scarce~\cite{Ma2019,Gianfrate2020, Lai2021,Li2024, Du2021}. The most favorable systems for facilitating the interpretation of measurements are  semimetals with low-energy band structures that are universally determined by a few parameters within a $k\cdot p$ expansion. The most well-studied instances are the relativistic dispersion of graphene in two dimensions (2D) -- characterized by a Fermi surface with quantized Berry phase~\cite{Zhang2005} -- and that of Weyl semimetals in 3D -- characterized by a Fermi surface with quantized Chern number, the chirality of the Weyl fermion~\cite{RevModPhys.90.015001}. The Berry curvature of Bloch states diverges as one approaches the Weyl node, adding another reason why quantum geometric quantities are favorably studied in topological semimetals~\cite{Zhang2016}. Multifold fermion semimetals present a natural extension beyond these elementary examples. They are defined by hosting symmetry-enforced band degeneracies of three or more bands at high-symmetry points in the Brillouin zone~\cite{bradlyn2016beyond}. For each band's Fermi surface, a Chern number can be computed, allowing one to assign a chirality to multifold fermions as well. Furthermore, their effective  $k\cdot p$ Hamiltonian maps to that of a spin-1 or spin-3/2 object, for the three- and four-band case, respectively, (with momentum relative to the high-symmetry point taking the role of magnetic field) providing a simple classification of multifold fermions.
\begin{figure*}
    \centering
    \includegraphics[width=\linewidth]{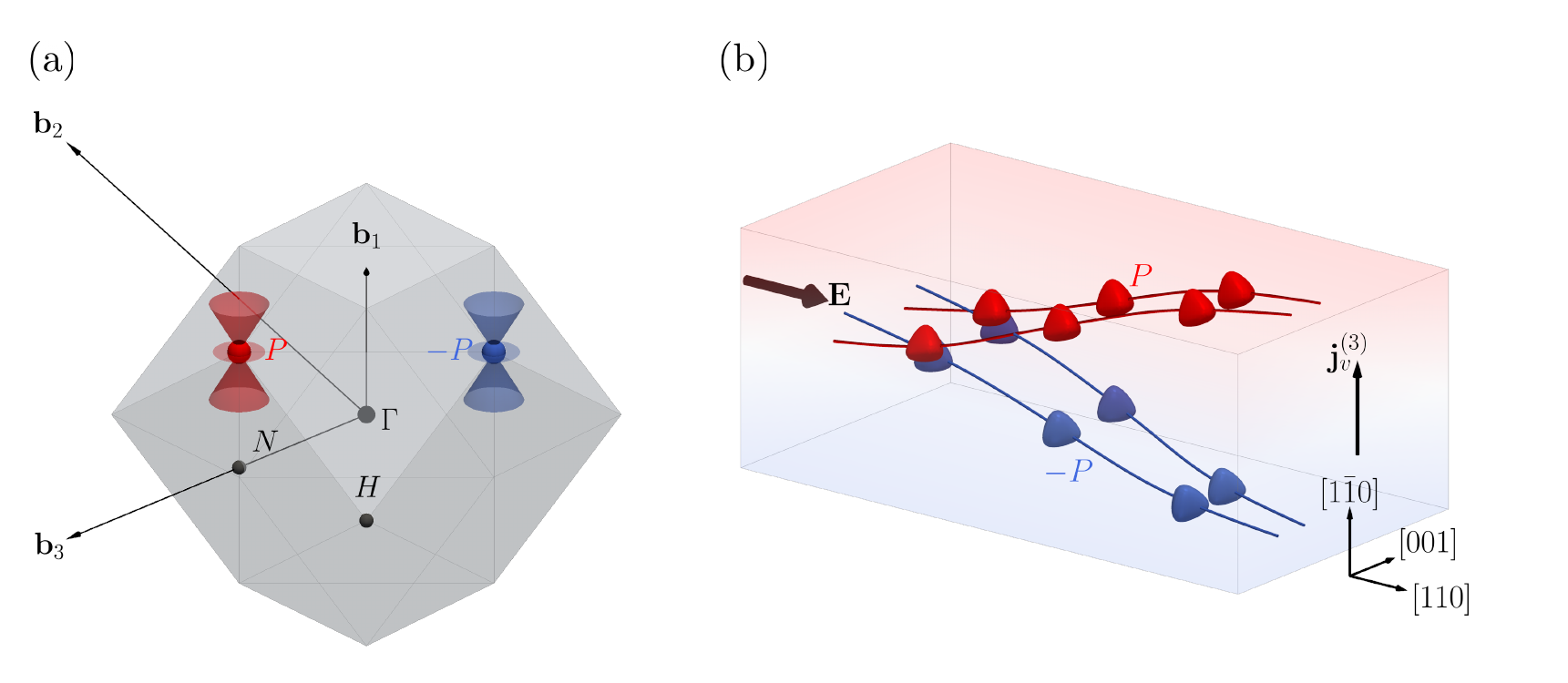}
    \caption{(a) The Brillouin zone of space group $I4_132$, this symmetry group allows for three-fold crossings at the $P$ points, forming two valleys with spin-1 fermions as low energy degrees of freedom. (b) When an electric field is applied to the material along $[110]$, a non-linear third order transverse valley current is formed along $[1\bar{1}0]$, leading to accumulation of electrons from each pocket on the top and bottom surfaces of the material. The direction and magnitude of the orbital magnetic moment accumulation can be tuned by the chemical potential.}
    \label{fig:coverfig}
\end{figure*}

The objective of this work is to establish a framework for relating bulk non-linear conductivity measurements in multifold fermion systems~\cite{bradlyn2016beyond, robredo2024multifoldtopologicalsemimetals} to their quantum geometry. We focus on time-reversal symmetric crystals and the DC (low frequency) regime. Our calculations are within Boltzmann transport theory in the constant relaxation time approximation exposing both Drude (originating from the band dispersion) as well as quantum geometric contributions up to third order in the electric field (Sec.~\ref{sec:formalism}). Experimentally, distinguishing intrinsic versus extrinsic as well as quantum geometric versus Drude contributions is a key challenge. We furthermore provide relaxation time scaling arguments, as well as symmetry arguments about the structure of the conductivity tensors that are valid beyond the approximations of semiclassical transport (Sec.~\ref{sec: symmetry}). In most crystalline space groups, multifold fermions are located at time-reversal symmetric momenta, in which case time-reversal symmetry imposes a vanishing of several response coefficients. However, in select space groups, this is not the case (see Fig.~\ref{fig:coverfig}a), opening the possibility of multifold fermions with oppositely oriented valley-dependent responses and with this to non-linear valley-tronics~\cite{Fan2024}. We illustrate the two options with an in-depth discussion of generalized crossings in an idealized limit with SO(3) rotation symmetry, and in three-fold crossings arising in materials with space group symmetry $I4_132$ (Sec.~\ref{sec:kpmodels}) where a transverse third order valley current can arise (see Fig.~\ref{fig:coverfig}b). Finally, we discuss an experiment on PdGa (space group $P2_13$) in light of our results~\cite{Dixit2025}.

\section{Semi-classical formalism}\label{sec:formalism}
We follow the arguments in Refs.~\onlinecite{sodemann2015quantum,liu2025quantum, jia2024equivalence, jiang2025revealing} and provide a summary of results on non-linear transport fully consistent up to third order in the electric field strength within the semi-classical Boltzmann formalism. The Boltzmann equation for a species of electrons (i.e. within a single, isolated \enquote{active} band, labelled \enquote{0}) subject to an external electric field $\textbf{E}(t)$ is given by
\begin{equation}
    \frac{d\textbf{r}}{dt} \cdot \nabla_\textbf{r} f + \frac{d\textbf{k}}{dt} \cdot \nabla_\textbf{k} f + \frac{\partial f}{\partial t} = \bigg(\frac{\partial f}{\partial t}\bigg)_\text{coll},
\end{equation}
where $f(\textbf{r},\textbf{k},t)$ is the probability density function for the electrons in position-momentum phase space $(\textbf{r}(t),\textbf{k}(t))$. This equation describes the relaxation of the electronic distribution in phase space due to all possible scattering channels, captured by the collision integral $\big(\frac{\partial f}{\partial t}\big)_\text{coll}$. These include but are not limited to scattering off impurities of various kinds as well as electron-phonon and electron-electron scattering. It is in general very difficult to obtain an accurate microscopic collision integral, so some approximations on its form are usually made. The crudest approximation is the relaxation-time approximation (RTA), where we approximate the collision term by $\big(\frac{\partial f}{\partial t}\big)_\text{coll} \approx \frac{f_0-f}{\tau}$ with $f_0$ being the equilibrium distribution function and $\tau$ being a phenomenological parameter which can depend, for instance, on crystalline momentum, temperature and band/Fermi pocket. Although crude, the RTA works remarkably well in capturing transport properties. The approximation cannot capture interband-coherent effects, and it is only justified in the diffusive regime $\tau \ll L_\text{sample}/v_\text{F}$, where $L_\text{sample}$ is the length of the sample and $v_\text{F}$ is the Fermi velocity averaged around the Fermi surface. 

The dynamics of the phase space variables $(\textbf{r}, \textbf{k})$ are described by the semi-classical equations of motion which, in the absence of an external magnetic field, take the following form
\begin{align}\label{semiclassicaleom}
\begin{cases}
    \frac{d\textbf{r}}{dt} = \frac{1}{\hbar}\nabla_\textbf{k}\bar\varepsilon +\frac{e}{\hbar}\textbf{E} \times \bar{\boldsymbol{\Omega}} \\
    \frac{d\textbf{k}}{dt} = -\frac{e}{\hbar}\textbf{E}
\end{cases}.
\end{align}
Here $\bar \varepsilon$ and $\bar{\boldsymbol{\Omega}}$ are the band energy dispersion and Berry curvature of the active band respectively, including corrections coming from the electric field~\cite{gao2014field}. More explicitly, we can expand these two quantities in orders of the electric field (denoted by the superscript in parenthesis):
\begin{equation}
\begin{cases}
    \bar \varepsilon =\varepsilon^{(0)}+\varepsilon^{(1)}+\varepsilon^{(2)}+...\\
    \bar{\boldsymbol{\Omega}} =\boldsymbol{\Omega}^{(0)}+\boldsymbol{\Omega}^{(1)}+\boldsymbol{\Omega}^{(2)}+...
\end{cases}.
\end{equation}
The generalized semiclassical equations of motion in Eq.~\eqref{semiclassicaleom} describe the evolution of an electron wave-packet centered at $(\textbf{r},\textbf{k})$ formed from superposing various Bloch states, and can be perturbatively expanded in powers of the electric field. They are therefore accurate in the weak-field, low-frequency regime $\hbar \omega \ll \Delta$ where $\omega$ is the frequency of the external electric field and $\Delta$ the characteristic energy scale of the band. We therefore cannot capture multi-band resonance effects, which would instead require a full quantum mechanical treatment (see Refs.~\onlinecite{parker2019diagrammatic,jia2024equivalence,ulrich2025quantum,du2021quantum,jiang2025revealing}). We are also limited to cases of non-degenerate active bands, if the bands are degenerate somewhere on the Fermi surface, then many of the quantum geometric quantities that we define in the rest of the paper become singular. We will focus on bulk transport for which we can consider a translationally invariant system where $\nabla_\textbf{r} f=0$. The Boltzmann equation then takes the form
\begin{equation}
    \frac{\partial f(\textbf{k},t)}{\partial t} -\frac{e}{\hbar}\textbf{E}(t)\cdot \nabla_\textbf{k} f(\textbf{k},t) = \frac{f_0(\textbf{k})-f(\textbf{k},t)}{\tau}.
\end{equation}

For the sake of brevity, we will adopt the following shorthand notation $\frac{\partial}{\partial t} \equiv \partial_t$ and $\frac{\partial}{\partial k_\alpha} \equiv \partial_\alpha$ and use index-summed notation unless otherwise specified. Let us decompose $f$ into a power series in the electric field intensity $E$ so $f-f_0={f}_1+{f}_2+{f}_3+...$ where $f_n$ scales as $E^n$. By equating terms in powers of the electric field $E$, we obtain a recursive system of differential equations:
\begin{equation*}
    (1+\tau\partial_t) f_n - \frac{e\tau}{\hbar} E_\alpha \partial_\alpha f_{n-1} = 0.
\end{equation*}
These can be solved by inverting the kernel on the first term, which we write formally as:
\begin{equation}\label{eq:frecursive}
    f_n= \bigg(\frac{e\tau}{\hbar} \frac{1}{1+\tau \partial_t}E_\alpha \partial_\alpha\bigg)^n f_{0}.
\end{equation}
We give the explicit form of $f_1$, $f_2$ and $f_3$ for a monochromatic AC electric field in App.~\ref{app:ACdistr}, but  will otherwise work in the limit of $\omega \tau \ll 1$ where $f_n \approx (\frac{e \tau}{\hbar} \textbf{E}\cdot \nabla_\textbf{k})^nf_0$. Moreover, we would like to remark that $f_0$ is the equilibrium Fermi-Dirac distribution evaluated at the bare band energies $\varepsilon^{(0)}$. Within the relaxation time approximation, this corresponds to assuming that scattering processes relax the electrons towards the equilibrium state of the field-free crystal, while field-induced corrections enter the dynamics only through the semi-classical equations of motion. Other works~\cite{qiang2025clarification, xiao2024definition, xiang2023third, gao2014field} instead propose to use the equilibrium distribution function evaluated at the band-corrected energies $\tilde{f}_0(\textbf{k}) = \frac{1}{e^{\beta(\tilde{\varepsilon}(\textbf{k})-\varepsilon_\text{F})}+1}$, obtaining different expressions for the conductivity, and in particular no intrinsic longitudinal response. The bare-energy prescription is supported by the demonstrated agreement between semiclassical and quantum response theory \cite{jia2024equivalence, ulrich2025quantum}. Nevertheless, a more complete treatment of the collision integral from a microscopic standpoint is necessary in order to determine which distribution the system relaxes to.

Having found the non-equilibrium distribution to arbitrary order in the electric field, we proceed to extract the current densities. The total current density is found by integrating the distribution-weighted velocity:
\begin{align}
    j_\alpha(t) &= \int_k-e v_\alpha(\textbf{k},t) f(\textbf{k},t) \nonumber \\
    &=-\frac{e}{\hbar}\int_k \partial_\alpha \bar{\varepsilon}(\textbf{k})  f(\textbf{k},t)-\frac{e^2}{\hbar}\int_k \epsilon_{\alpha \beta \gamma} E_\beta(t)\bar\Omega_\gamma(\textbf{k}) f(\textbf{k},t),
\end{align}
where we defined the normalized phase space integral $\int_k \equiv \int \frac{d^dk}{(2\pi)^d}$. We will thus obtain two types of terms at each order, a Drude-like term involving $\partial_\alpha \bar \varepsilon$ and a $k$-space moment of the Berry curvature $\bar\Omega_\beta$. We proceed to define
\begin{equation}\label{eq:Efieldcorrection}
\begin{cases}
\bar\varepsilon^{(m)}(\textbf{k}) = D^{(m)}_{\alpha_1...\alpha_m}(\textbf{k}) E_{\alpha_1} ... E_{\alpha_m}, \\ \bar{\Omega}_\alpha^{(m)}(\textbf{k}) = T^{(m)}_{\alpha;\alpha_1...\alpha_m}(\textbf{k}) E_{\alpha_1} ... E_{\alpha_m}
\end{cases}
\end{equation}
to be the $m$th order electric field induced corrections to the band energy and Berry curvature of the wave-packet respectively. By definition, we then have that $D^{(0)}(\textbf{k}) = \varepsilon^{(0)}(\textbf{k})$ and $T_\alpha^{(0)}(\textbf{k}) = \Omega^{(0)}(\textbf{k})$. The $n$th order current response is then found to be:
\begin{align}\label{general_current}
    j_\alpha^{(n)}(t) = &-\frac{e}{\hbar}\sum_{m=0}^n\int_k \partial_\alpha D_{\alpha_1...\alpha_m}^{(m)}E_{\alpha_1} ... E_{\alpha_m} f_{n-m}\nonumber \\
    &-\frac{e^2}{\hbar} \sum_{m=0}^{n-1}\int_k \epsilon_{\alpha \beta \gamma} T_{\gamma;\alpha_1...\alpha_m}^{(m)} E_\beta E_{\alpha_1} ... E_{\alpha_m} f_{n-(m+1)}.
\end{align}
This current can then be further decomposed into harmonics $n\omega,(n-2)\omega,...$. The electric field induced corrections to $\varepsilon$ and $\boldsymbol{\Omega}$ can be computed following Refs.~\onlinecite{gao2014field,xiang2023third} by deriving the Lagrangian for a semi-classical wave-packet centered at $(\textbf{r}_c,\textbf{k}_c)$ with an electric field induced positional shift, or following Refs.~\onlinecite{kaplan2024unification,fang2024quantum} using a Luttinger-Kohn formalism. These two approaches give, rather inconsistently, different corrections to the energy. This discrepancy, as explained in Refs.~\onlinecite{xiang2023third,jia2024equivalence}, arises from a difference between the energy of the wave-packet obtained by perturbation theory, and the final energy which enters the semi-classical equations of motion. Indeed, the wave-packet energy can be expanded as 
\begin{equation}\label{eq:energycorr1}
\bar \varepsilon = \varepsilon^{(0)} + e\textbf{E}\cdot \textbf{r}_c^{(1)} +\tilde{\varepsilon}^{(2)}+O(E^3),
\end{equation}
where $\textbf{r}_c^{(1)}=\textbf{r}_c^{(0)} + \boldsymbol{\mathcal{A}}^{(1)}$ is the wave-packet position to first order in the electric field given by ${\mathcal{A}}^{(1)}_\alpha=eG_{\alpha \beta}E_\beta$,  $\tilde{\varepsilon}_0^{(2)} = -\frac{1}{2}e^2G_{\alpha \beta}E_\alpha E_\beta$ is the field-induced band energy correction, and $G_{\alpha \beta} = 2\text{Re}\Big[\sum_{n\neq0}\frac{A^\alpha_{0n}A^\beta_{n0}}{\varepsilon_0-\varepsilon_n}\Big]$ is the Berry connection polarizability (BCP) tensor~\cite{gao2014field}. Note crucially that $\tilde{\varepsilon}^{(2)}$ is not the second order energy correction obtained in perturbation theory; the correct perturbative expansion of the energy would instead be expanded as
\begin{equation}\label{eq:energycorr2}
    \bar \varepsilon = \varepsilon^{(0)} + e\textbf{E}\cdot \textbf{r}_c^{(0)} +(\tilde{\varepsilon}^{(2)}+e\textbf{E}\cdot \boldsymbol{\mathcal{A}}^{(1)})+O(E^3).
\end{equation}
Hence, the second order energy correction is $\varepsilon^{(2)}=\tilde{\varepsilon}^{(2)}+e\textbf{E}\cdot \boldsymbol{\mathcal{A}}^{(1)} = \frac{1}{2}e^2G_{\alpha\beta} E_\alpha E_\beta$, which matches with the energy correction obtained from perturbation theory~\cite{kaplan2024unification}. In the semi-classical formalism, however, one must then derive the equations of motion for the wave-packet from its Lagrangian. These turn out to be~\cite{gao2014field}
\begin{equation*}
    \frac{d\textbf{r}_c}{dt} = \frac{1}{\hbar}\nabla_{\textbf{k}_c} \bar\varepsilon + \frac{e}{\hbar}\textbf{E}\times \bar{\boldsymbol{\Omega}}, \ \frac{d\textbf{k}_c}{dt} = -\frac{e}{\hbar}\textbf{E},
\end{equation*}
and since $\textbf{k}_c$ and $\textbf{r}_c$ are treated as independent phase-space variables, we see from Eq.~\eqref{eq:energycorr1} that the energy entering the equations of motion up to second order in $E$ does not contain terms dependent on $\textbf{r}_c$:
\begin{equation*}
    \bar \varepsilon^\text{EOM} =  \varepsilon^{(0)}  +\tilde{\varepsilon}^{(2)}+O(E^3) =\varepsilon^{(0)}  +({\varepsilon}^{(2)}-e\textbf{E}\cdot \boldsymbol{\mathcal{A}}^{(1)})+O(E^3).
\end{equation*} 
In a semiclassical formalism, $D^{(m)}_{\alpha_1...\alpha_m}$ should be viewed as encapsulating the $m$th order corrections to the energy entering the equations of motion, rather than the perturbative correction to the energy \cite{gao2014field}. Hence, the third order correction to the energy entering the equations of motion, $\tilde{\varepsilon}^{(3)}$, can be obtained from the total third order correction to the energy $\varepsilon^{(3)}$ by subtracting the second order positional shift: $\tilde{\varepsilon}^{(3)} = \varepsilon^{(3)}-e\textbf{E}\cdot \boldsymbol{\mathcal{A}}^{(2)}$. The third order perturbative energy correction obtained in Ref.~\onlinecite{xiang2023third} is
\begin{align}\label{energycorr3}
\varepsilon^{(3)}=e^3 \operatorname{Re}\Bigg[&\sum_{n \neq 0}\sum_{m \neq 0,n} \frac{A^\alpha_{0n}A^\beta_{nm}A^\gamma_{m0}}{(\varepsilon_0-\varepsilon_n)(\varepsilon_0-\varepsilon_m)} \nonumber\\
+&\sum_{n\neq 0} \frac{A^\alpha_{0n}(i\partial_\beta+A_n^\beta-A_{0}^\beta) A^\gamma_{n0}}{(\varepsilon_{0}-\varepsilon_n)^2}\Bigg] E_\alpha E_\beta E_\gamma,
\end{align}
where $A^\alpha_{nm} = i\braket{u_n|\partial_\alpha u_m}$ is the interband Berry connection. We re-arranged the result so that it closely mirrors the correction found in Ref.~\onlinecite{fang2024quantum}, but made gauge-invariant thanks to the covariant derivative in the second term. Indeed under a $U(1)$ gauge transformation $\ket{u_n} \mapsto e^{i\chi_n} \ket{u_n}$, the Berry connection transforms as $A_{nm}^\alpha \mapsto e^{i(\chi_m-\chi_n)} A_{nm}^\alpha-\delta_{nm}\partial_\alpha \chi_n$. Also, the quantity in brackets is implicitly symmetrized in $\alpha,\beta,\gamma$, as it is contracted with $E_\alpha E_\beta E_\gamma$. Substituting these into Eq.~\eqref{energycorr3}, one readily finds that it is gauge invariant. The second order positional shift was computed in Ref.~\onlinecite{xiang2023third} to be
\begin{align}
\mathcal{A}_\alpha^{(2)}
= e^2\operatorname{Re}\bigg[&\sum_{n}
\Bigl(
  2A_{0}^{\beta} M_{0n}^{\alpha} M_{n0}^{\gamma}
  + A_{0}^{\alpha} M_{0n}^{\beta} M_{n0}^{\gamma}
\Bigr)
\nonumber \\[4pt]
\quad -&\sum_{n}\sum_{m}
\Bigl(
  2M_{0n}^{\alpha} A_{nm}^{\beta} M_{m0}^{\gamma}
  + M_{0n}^{\beta} A_{nm}^{\alpha} M_{m0}^{\gamma}
\Bigr)
\nonumber \\[4pt]
\quad -&\sum_{n}
\Bigl(
  M_{0n}^{\beta} i\partial_{\alpha} M_{n0}^{\gamma}
  + M_{0n}^{\alpha} i\partial_{\beta} M_{n0}^{\gamma}
  - M_{n0}^{\gamma} i\partial_{\beta} M_{0n}^{\alpha}
\Bigr)\Bigg]
\nonumber\\[4pt]
& \times E_{\beta}E_{\gamma},
\end{align}
where $M_{nm}^\alpha = \frac{A_{nm}}{\varepsilon_m-\varepsilon_n}(1-\delta_{nm})$. Again one can show that the above is gauge invariant. Furthermore, it is also possible to show that $e\textbf{E}\cdot \boldsymbol{\mathcal{A}}^{(2)} = 3\varepsilon^{(3)}$, and consequently $\tilde \varepsilon^{(3)} = -\frac{2}{3}e \textbf{E}\cdot \mathcal{A}^{(2)}$. This relation holds more generally to any order:
\begin{equation}\label{HFeq}
    \varepsilon^{(n)} = \frac{1}{n}e\textbf{E}\cdot \boldsymbol{\mathcal{A}}^{(n-1)} \implies \tilde \varepsilon^{(n)} = -\frac{n-1}{n}e \textbf{E}\cdot \boldsymbol{\mathcal{A}}^{(n-1)},
 \end{equation}
which can be readily proven using the Hellmann-Feynman theorem (see App.~\ref{app:HF}). This implies that the conductivity tensor can be expressed entirely in terms of the positional shifts. We rearrange Eq.~\eqref{general_current} in orders of the corrections to the distribution function:
\begin{align*}
    j_\alpha^{(n)}(t) = &-\frac{e}{\hbar}\int_k \partial_\alpha \varepsilon^{(0)}f_{n}(\textbf{k},t) 
    \nonumber \\
    &-\frac{e}{\hbar} \sum_{m=1}^{n}\int_k (\partial_\alpha D^{(m)}_{\alpha_1...\alpha_m}+e\epsilon_{\alpha \alpha_m \gamma} T_{\gamma;\alpha_1...\alpha_{m-1}}^{(m-1)})
    \nonumber\\
    & \hspace{2cm}\times E_{\alpha_1}(t) ... E_{\alpha_{m}}(t) f_{n-m}(\textbf{k},t).
\end{align*}
The terms on the second line are further simplified by defining $\mathcal{A}_\alpha^{(n)} = e^n\mathcal{G}^{(n)}_{\alpha;{\alpha_1}...{\alpha_n}} E_{\alpha_1} ... E_{\alpha_n}$ (so that $\mathcal{G}^{(0)}_\alpha =  A^{(0)}_\alpha$ is the bare Berry connection). We then get:
\begin{align}\label{finalcurrent}
    j_\alpha^{(n)}(t) = &-\frac{e}{\hbar}\int_k \partial_\alpha \varepsilon^{(0)}f_{n}(\textbf{k},t) 
    \nonumber \\
    &-\frac{e}{\hbar} \sum_{m=1}^{n}\int_k \frac{e^m}{m}\bigg(\partial_\alpha \mathcal{G}^{(m-1)}_{\alpha_m;\alpha_1...\alpha_{m-1}}-\sum_{j=1}^m\partial_{\alpha_j} \mathcal{G}^{(m-1)}_{\alpha;\{\alpha_i\}\setminus\alpha_j}\bigg)
    \nonumber\\
    & \hspace{2cm} \times E_{\alpha_1}(t) ... E_{\alpha_{m}}(t) f_{n-m}(\textbf{k},t),
\end{align}
where $\{\alpha_i\}\setminus\alpha_j$ indicates that this string of indices is $\alpha_1...\alpha_m$ excluding $\alpha_j$. 

Using Eq.~\eqref{finalcurrent}, we compute the non-linear response of a system subject to an external electric field up to third order. At first order we have the first-harmonic response
\begin{align}\label{firstorder}
   j^{(1)}_\alpha(t) &= -\frac{e}{\hbar} \int_k \partial_\alpha \varepsilon^{(0)} f_1+ \frac{e^2}{\hbar}\int_k \epsilon_{\alpha \beta \gamma} \Omega_\beta {E}_\gamma f_0 .
\end{align}
The first term is the standard linear Drude current $j^{(1),\text{D}}$, while the second is the anomalous Hall current $j^{(1),\text{BC}}$. Note that for systems with time-reversal symmetry (TRS), the second term vanishes because the integrand is antisymmetric in \textbf{k}. At second order, we instead have 
\begin{align}\label{secondorder}
    {j}^{(2)}_\alpha(t) =&-\frac{e}{\hbar}\int_k \partial_\alpha \varepsilon^{(0)}f_2 + \frac{e^2}{\hbar}\int_k \epsilon_{\alpha \beta \gamma} \Omega_\beta {E}_\gamma f_1
    \nonumber\\
    &-\frac{e^3}{2\hbar}\int_k (\partial_\alpha \mathcal{G}^{(1)}_{\alpha_1 \alpha_2} - \partial_{\alpha_1} \mathcal{G}^{(1)}_{\alpha \alpha_2} -\partial_{\alpha_2} \mathcal{G}^{(1)}_{\alpha \alpha_1}) E_{\alpha_1} E_{\alpha_2} f_0.
\end{align}
The first term is the second order Drude term $j^{(2),\text{D}}$, the second term is the Berry curvature dipole $j^{(2),\text{BCD}}$ and the third term is the Berry connection polarizability term $j^{(2),\text{BCP}}$ coming from the first order positional shift.  For systems with TRS, $j^{(2),\text{D}}$ vanishes because it involves an odd number of $k$-derivatives of energy. Similarly, $j^{(2),\text{BCP}}$ vanishes if TRS is present. Thus, we expect the second-harmonic response in TR symmetric systems to arise purely from $j^{(2),\text{BCD}}$. Finally, at third order we find 
\begin{align}\label{thirdorder}
    {j}^{(3)}_\alpha(t) =&-\frac{e}{\hbar}\int_k \partial_\alpha \varepsilon^{(0)} f_3+\frac{e^2}{\hbar}\int_k \epsilon_{\alpha \beta \gamma} \Omega_\beta E_\gamma f_2
    \nonumber \\
    &-\frac{e^3}{2\hbar}\int_k (\partial_\alpha \mathcal{G}^{(1)}_{\alpha_1 \alpha_2} - \partial_{\alpha_1} \mathcal{G}^{(1)}_{\alpha \alpha_2} -\partial_{\alpha_2} \mathcal{G}^{(1)}_{\alpha \alpha_1}) E_{\alpha_1} E_{\alpha_2} f_1
    \nonumber \\
    &-\frac{e^4}{3\hbar}\int_k (\partial_\alpha \mathcal{G}^{(2)}_{\alpha_1 \alpha_2 \alpha_3} - \partial_{\alpha_1} \mathcal{G}^{(2)}_{\alpha \alpha_2 \alpha_3} -\partial_{\alpha_2} \mathcal{G}^{(2)}_{\alpha \alpha_1 \alpha_3} \nonumber\\
    &\hspace{1.5cm}-\partial_{\alpha_3} \mathcal{G}^{(2)}_{\alpha \alpha_1 \alpha_2} ) E_{\alpha_1} E_{\alpha_2} E_{\alpha_3} f_0.
\end{align}
The first term is the third order Drude current $j^{(3),\text{D}}$, the second term is the Berry curvature quadrupole $j^{(3),\text{BCQ}}$, the third term is the Berry connection polarizability dipole $j^{(3),\text{BCPD}}$, and finally the fourth term is the second Berry connection polarizability $j^{(3),\text{SBCP}}$. In systems with TRS, the second and fourth terms vanish identically. 

Finally, we wish to remark that all these quantities can be expressed as Fermi-surface integrals when evaluated at zero temperature (see App.~\ref{app:condFS} for details).

\section{Symmetry analysis}\label{sec: symmetry}
The symmetries of a material will impose several constraints on the general form of the conductivity tensor. In the case of a crystalline material, the relevant symmetries are contained in its point group (or magnetic point group for crystals breaking TRS). We must thus identify under which representation of the point group the respective tensors transform. Assuming a constant relaxation time $\tau$, Eq.~\eqref{eq:frecursive} can be evaluated exactly. Inserting the solution $f_n = \big(\frac{e\tau}{\hbar} E_\alpha \partial_\alpha\big)^n f_0$ into Eq.~\eqref{finalcurrent}, one obtains
\begin{align}\label{eq:finalcurrent}
    j_\alpha^{(n)}(t) = &-\frac{e}{\hbar}\bigg(-\frac{e\tau}{\hbar}E_{\beta_i}\bigg)^{n}\int_k f_0\partial_\alpha \partial_{\beta_1} ... \partial_{\beta_n}\varepsilon^{(0)} 
    \nonumber \nonumber\\
    &-\frac{e}{\hbar} \sum_{m=1}^{n} \frac{e^m}{m}\bigg(-\frac{e\tau}{\hbar}E_{\beta_i}\bigg)^{n-m} E_{\alpha_1} ... E_{\alpha_{m}}  \nonumber\\
    &\times\int_kf_0\partial_{\beta_1}...\partial_{\beta_{n-m}} \bigg(\partial_\alpha \mathcal{G}^{(m-1)}_{\alpha_m;\alpha_1...\alpha_{m-1}}-\sum_{j=1}^m\partial_{\alpha_j} \mathcal{G}^{(m-1)}_{\alpha;\{\alpha_i\}\setminus\alpha_j}\bigg) 
\end{align}
after integrating by parts several times. We thus identify two types of tensors:
\begin{subequations}\label{eq:tensors}
\begin{align}
    &\mathcal{V}^{(n)}_{\alpha,\{\beta_i\}} = \int_k f_0\partial_\alpha \partial_{\beta_1} ... \partial_{\beta_n}\varepsilon^{(0)},\label{eq:tensors1} \\
    &\mathcal{U}^{(n,m)}_{\alpha,\{\alpha_i\},\{\beta_i\}} = \int_kf_0\partial_{\beta_1}...\partial_{\beta_{n-m}} \bigg(\partial_\alpha \mathcal{G}^{(m-1)}_{\alpha_m;\alpha_1...\alpha_{m-1}}-\sum_{j=1}^m\partial_{\alpha_j} \mathcal{G}^{(m-1)}_{\alpha;\{\alpha_i\}\setminus\alpha_j}\bigg).\label{eq:tensors2}
\end{align}
\end{subequations}
We are interested in how these tensors transform under the (magnetic) point group symmetries. In general, a Cartesian tensor (e.g.~the group velocity) will transform as
\begin{equation}
    T'_{\alpha'_1...\alpha'_n} = 
    \begin{cases}
    R^g_{\alpha'_1\alpha_1}...R^g_{\alpha'_n\alpha_n} T_{\alpha_1...\alpha_n} &\text{ TR even}\\
    \text{sgn}({\mathcal{K}^g}) R^g_{\alpha'_1\alpha_1}...R^g_{\alpha'_n\alpha_n} T_{\alpha_1...\alpha_n} &\text{ TR odd}
    \end{cases}
\end{equation}
under a magnetic point group operation $g$ where $R^g$ is the unitary component acting in real space and $\mathcal{K}^g$ is the anti-unitary component. Here we defined $\text{sgn}({\mathcal{K}^g})$ as $+1$ if $g$ is unitary and $-1$ if $g$ is anti-unitary. On the other hand, a pseudo-tensor (e.g.~the Berry curvature) will transform as
\begin{align}
    \Omega'_{\alpha'_1...\alpha'_n} = 
    \begin{cases}
    \det R^g \cdot R^g_{\alpha'_1\alpha_1}...R^g_{\alpha'_n\alpha_n} \Omega_{\alpha_1...\alpha_n} & \text{  TR even}\\
    \text{sgn}({\mathcal{K}^g}) \det R^g \cdot R^g_{\alpha'_1\alpha_1}...R^g_{\alpha'_n\alpha_n} \Omega_{\alpha_1...\alpha_n} & \text{  TR odd}
    \end{cases}.
\end{align}
In our case we have that $\mathcal{V}^{(n)}_{\alpha, \{\beta_i\}}$ is a Cartesian tensor with TR parity $(-1)^{n+1}$, and that $\mathcal{U}^{(n,m)}_{\alpha,\{\alpha_i\},\{\beta_i\}}$ is a Cartesian tensor with TR parity $(-1)^{n-m+1}$. In addition to the constraints imposed by the point-group symmetry, it follows from Eq.~\eqref{eq:finalcurrent} that we are only interested in the components of $\mathcal{V}^{(m)}_{\alpha, \{\beta_i\}}$ that are symmetric under permutation of indices in the set $\{\beta_i\}$. Similarly, we are only interested in the components of $\mathcal{U}^{(n,m)}_{\alpha,\{\alpha_i\},\{\beta_i\}}$ that are symmetric under permutation of indices in $\{\alpha_i, \beta_i\}$. 

\begin{table*}[t!]
    \centering
    \begin{tabular}{c|c|c|c|c|c|c|c}
        Point group & $T$ & $T_h$ & $T_d$ & $O$ & $O_h$ & $D_{4h}$ & SO(3)\\[3pt]
        \hline \hline
        1st order & $a_1=a_2$ & $a_1=a_2$ & $a_1=a_2$ & $a_1=a_2$ & $a_1=a_2$ & $-$ & $a_1=a_2$ \\[3pt]
        \hline
        2nd order & $-$ & $b_1=0$ & $-$ & $b_1=0$ & $b_1=0$ & $b_1=0$ & $b_1=0$ \\[3pt]
        \hline
        3rd order 
        & $\begin{array}{c}
             c_3 = c_1\\
             c_4 = c_2\\
             c_6 = c_5
           \end{array}$
        & $\begin{array}{c}
             c_3 = c_1\\
             c_4 = c_2\\
             c_6 = c_5
           \end{array}$
        & $\begin{array}{c}
             c_3 = c_4 = c_1 = c_2\\
             c_6 = c_5
           \end{array}$
        & $\begin{array}{c}
             c_3 = c_4 = c_1 = c_2\\
             c_6 = c_5
           \end{array}$
        & $\begin{array}{c}
             c_3 = c_4 = c_1 = c_2\\
             c_6 = c_5
           \end{array}$
        & $c_3 = c_4$ & 
         $c_i$ all equal
    \end{tabular}
    \caption{Symmetry-enforced constraints on the first, second and third order conductivity matrix elements for the point groups hosting multifold crossings. The definition of $a_i, b_i, c_i$ can be found in Eq.~\eqref{jEgeneral1} and  Eq.~\eqref{jEgeneral2}.}
    \label{tab:symmetry_table}
\end{table*}

We will focus on non-magnetic materials hosting multifold crossings, which only occur at high symmetry momenta with little co-group $T, T_h, T_d, O, O_h$ or $D_{4h}$ \footnote{note that these are all cubic except for $D_{4h}$ which is tetragonal. For $D_{4h}$, we take the $z$-axis to be the $C_4$ axis.} (see App.~\ref{app:multifold_tables} for more details on the classification of multifold semimetals). For all but the last point group, the most general form of the constitutive relations are the following:
\begin{equation}\label{jEgeneral1}
\begin{cases}
    j_x = a_1 E_x + b_1 E_y E_z + (c_1 E_y^2 + c_2 E_z^2 + c_5 E_x^2)E_x\\ 
    j_y = a_1 E_y + b_1 E_x E_z + (c_1 E_z^2 + c_2 E_x^2 + c_5 E_y^2)E_y \\ 
    j_z = a_2 E_z + b_1 E_y E_x + (c_3 E_x^2 + c_4 E_y^2 + c_6 E_z^2)E_z
\end{cases},
\end{equation}
while for $D_{4h}$ the relations are:
\begin{equation}\label{jEgeneral2}
\begin{cases}
    j_x = a_1 E_x + b_1 E_y E_z + (c_1 E_z^2 + c_2 E_y^2 + c_5 E_x^2)E_x\\ 
    j_y = a_1 E_y + b_1 E_x E_z + (c_1 E_z^2 + c_2 E_x^2 + c_5 E_y^2)E_y \\ 
    j_z = a_2 E_z + b_1 E_y E_x + (c_3 E_x^2 + c_4 E_y^2 + c_6 E_z^2)E_z
\end{cases}.
\end{equation}
Note the indices on the first two terms in the cubic responses are different. For each point group, some further simplifications then occur, which we list in Table~\ref{tab:symmetry_table}. 

The microscopic expressions for the coefficients $a_1,a_2,b_1,c_1,c_2,c_3,c_4,c_5,c_6$ can be obtained by expanding Eq.~\eqref{firstorder}, Eq.~\eqref{secondorder} and Eq.~\eqref{thirdorder} and retaining the tensor component that is symmetric in the electric field indices. Finally, we note that these relations must be supplemented by the constraints imposed by the tensorial structure of individual quantum geometric tensors when considering their contributions independently. 

To close the section, we summarize the different responses we defined and some of their characteristic properties in Table~\ref{tab:summary}.

\begin{table*}
    \centering
    \small
    \renewcommand{\arraystretch}{1.2}
    \begin{tabular}{c l l c c}
        \toprule
        Order
        & Contribution
        & Geometric quantity
        & $\tau$ scaling
        & TR parity
        \\
        \midrule

        1
        & Drude
        & Band dispersion $\partial_\alpha \varepsilon^{(0)}$
        & $\tau$
        & Even
        \\

        1
        & BC
        & Berry curvature $\Omega_\beta$
        & $\tau^{0}$
        & Odd
        \\

        \midrule

        2
        & Drude
        & Band dispersion $\partial_\alpha \partial_\beta \varepsilon^{(0)}$
        & $\tau^{2}$
        & Odd
        \\

        2
        & BCD
        & Berry curvature dipole $\partial_\alpha \Omega_\beta$
        & $\tau$
        & Even
        \\

        2
        & BCP
        & Berry connection polarizability $\mathcal{G}^{(1)}$
        & $\tau^{0}$
        & Odd
        \\

        \midrule

        3
        & Drude
        & Band dispersion $\partial_\alpha \partial_\beta \partial_\gamma \varepsilon^{(0)}$
        & $\tau^{3}$
        & Even
        \\

        3
        & BCQ
        & Berry curvature quadrupole $\partial_\alpha \partial_\beta \Omega_\gamma$
        & $\tau^{2}$
        & Odd
        \\

        3
        & BCPD 
        & Berry connection polarizability dipole $\partial_{\alpha}\mathcal{G}^{(1)}$
        & $\tau$
        & Even
        \\

        3
        & SBCP
        & Second Berry connection polarizability $\mathcal{G}^{(2)}$
        & $\tau^{0}$
        & Odd
        \\

        \bottomrule
    \end{tabular}
    \caption{Summary of the different contributions to the conductivity up to third order, the relevant geometric quantity,  the scaling of the contribution with the relaxation time $\tau$, and the contribution's time-reversal parity. }
    \label{tab:summary}
\end{table*}
\section{Case study: multifold fermions}\label{sec:kpmodels}
We will now apply the formalism developed in Sec.~\ref{sec:formalism} and Sec.~\ref{sec: symmetry} to investigate the low-frequency bulk transport in multifold semimetals. We will focus on three examples: (i) idealized limit with full SO(3) point symmetry, (ii) space group $I4_132$ which can host a three-fold crossing at $P$, which is \emph{not} a TRIM, (iii) space group $P2_13$ which can host a four-fold crossing at $\Gamma$ and a six-fold crossing at $R$, which are both time-reversal invariant momenta (TRIMs). Based on the work in Ref.~\onlinecite{bradlyn2016beyond}, we construct $k\cdot p$ models describing the low energy physics near each of these crossings and compute all the  contributions to the non-linear conductivity up to third order. We note that more elaborate symmetry-adapted $k \cdot p$ models incorporating couplings to the external fields directly have recently been developed in Ref.~\onlinecite{SatowYamakage2025}.

\subsection{Multifold fermions in the idealized limit}
We begin by considering the idealized isotropic limit of a multifold crossing hosting spin-$S$ fermions with the following linearized low-energy Hamiltonian:
\begin{equation}
    H(\textbf{k}) = \hbar v_\text{F} \textbf{k} \cdot \textbf{S},
\end{equation}
where $\textbf{S}=(S_x,S_y,S_z)$ are spin-S matrices. This Hamiltonian possesses time-reversal symmetry as well as SO(3) rotation symmetry and can thus be viewed as the idealized limit of an isotropic multifold crossing when one ignores anisotropies from the actual point group in the material. The energy bands are $\varepsilon_m^{(0)}=m\hbar v_\text{F} k$ where $m=-S,-S+1,...,S$ are the band indices and $k=|\textbf{k}|$. We set the chemical potential $\mu$ away from the node at zero energy and compute the integrands of the conductivities analytically for each band (except $m=0$ for integer $S$, which is a perfectly flat band). We provide detailed calculations in App.~\ref{app:multifold_kS}. At first order, we find that a Fermi pocket from the $m$th band contributes
\begin{align}\label{eq:kscond1}
    &\sigma_{\alpha \beta}^{(1),\text{D}} = \frac{e^2 \tau \mu^2}{6 \pi^2 \hbar^3 v_\text{F} |m|}\delta_{\alpha \beta}, \quad \sigma_{\alpha\beta}^{(1),\text{BC}} = 0,
\end{align}
where anomalous Hall response vanishes due to the presence of TRS. At second order, we find that all contributions vanish $\sigma_{\alpha\beta \gamma}^{(2)} = 0$ (the Drude and BCP terms vanish due to time-reversal symmetry, while the BCD term vanishes due to the rotation symmetry, see Tab.~\ref{tab:symmetry_table}). One must therefore go to third order to see the effects from quantum geometry. We find that only the Drude and BCP dipole terms contribute (as required by time-reversal symmetry):
\begin{align}
    \sigma^{(3),\text{D}}_{\alpha \beta \gamma \eta} = -\frac{e^4 \tau^3 v_\text{F} |m|}{15 \pi^2 \hbar^3} (\delta_{\alpha \beta} \delta_{\gamma \eta} + \delta_{\alpha \gamma} \delta_{\beta \eta} + \delta_{\alpha \eta} \delta_{\beta \gamma}),\\
    \sigma^{(3), \text{BCPD}}_{\alpha \beta \gamma \eta}  = \frac{e^4 \tau v_\text{F} |m|^3}{18 \pi^2 \hbar \mu^2} (\delta_{\alpha \beta} \delta_{\gamma \eta} + \delta_{\alpha \gamma} \delta_{\beta \eta} + \delta_{\alpha \eta} \delta_{\beta \gamma})\label{eq:kscond3},
\end{align}
while $\sigma^{(3), \text{BCQ}} = \sigma^{(3),\text{SBCP}} = 0$, again by TRS. Note that since the two contributions have opposite signs and different scaling in $\mu$, it is possible in principle to control the dominant mechanism for the third order current by tuning the chemical potential. We thus see that the first symmetry-allowed quantum geometric contribution to the idealized multifold fermion conductivity comes from the Berry connection polarizability dipole. 
\subsection{Valley current in space group $I4_132$}
The idealized model investigated in the previous section is not adequate to describe multifold crossings at non-TRIMs due to its time-reversal invariance. For crossings at non-TRIMs in time-reversal invariant systems, it is conceivable that a valley current may arise from those contributions to the conductivity that are odd under time-reversal. In App.~\ref{app:multifold_tables} we summarize for convenience the exhaustive classification of multifold fermions first presented in Ref.~\onlinecite{bradlyn2016beyond}. We note that it is possible to have multifold crossings at non-TRIMs in 6 different space groups (the crossing is at the P point in all these cases). However, in all but two cases (space groups $I2_13$ and $I4_132$), the point group symmetry (or its combination with time-reversal) forces some if not all of the energy bands to remain degenerate along certain high-symmetry directions, making them unsuitable for our semi-classical formalism. More generally, the resolution of multifold degeneracies into ordinary Weyl points under generic perturbations has recently been investigated using singularity theory in Ref.~\onlinecite{Franketal2025}. We will therefore just focus on the two cases where there are no symmetry-enforced band degeneracies on the Fermi surface and investigate the resulting valley currents. 

Space groups $I2_13$ and $I4_132$ can host three-fold nodal points at the $P$ point. As previously mentioned, this point is not a TRIM, so together with its time-reversed partner, they form two valleys, $P$ and $-P$, as shown in Fig.~\ref{fig:coverfig}a. The low energy physics near the $P$ valley is captured by the following  $k \cdot p$  model~\cite{bradlyn2016beyond}:
\begin{equation}\label{eq:kp199}
    H_P(\delta \textbf{k}) = \hbar v_\text{F} \begin{pmatrix}
    0 & e^{i\phi} \delta k_x & e^{-i\phi} \delta k_y\\
    e^{-i\phi} \delta k_x & 0 & e^{i\phi} \delta k_z\\
    e^{i\phi} \delta k_y & e^{-i\phi} \delta k_z & 0 
\end{pmatrix},
\end{equation}
where $\delta \textbf{k}$ is the momentum measured from the $P$ point and $v_\text{F},\phi$ are model parameters. The Hamiltonian in the opposite valley is obtained by time-reversal conjugation $H_{-P}(\delta \textbf{k}) = H_P^*(-\delta \textbf{k})$. Note that when $\phi=\pi/2$, the Hamiltonian simplifies to $H_P = \hbar v_\text{F} \delta \textbf{k}\cdot \textbf{S}$ where $\textbf{S}$ is the spin-1 operator, so we return to the idealized limit with full $\text{SO}(3)$ rotation symmetry discussed previously. We will tune away from this limit to exploit the non-TRIM nature of the $P$ point. Furthermore, we also note that point group symmetry of this Hamiltonian is $T$, which agrees with the little co-group at $P$ for the two space groups. However, the two-valley model possesses an additional symmetry, namely a four-fold rotation exchanging the two valleys: $R_{4z}H_P(C_{4z}\textbf{k}) R_{4z}^{-1}= H_{-P}(\textbf{k})$ where $C_{4z}(k_x,k_y,k_z) = (-k_y,k_x,k_z)$ and $R_{4z} = 1 \oplus -i\sigma_y$. This symmetry is present in space group $I4_132$ but not in $I2_13$. In the latter case it is an accidental symmetry arising from the linearization of the $k\cdot p$ model, and would thus be broken when keeping quadratic terms in the model. For the sake of simplicity, we will only consider the case of space group $I4_132$ where the linear $k\cdot p$ model is adequate. 

We set $\hbar v_\text{F} = 3~\text{eV \AA}$ (corresponding to $v_\text{F} \sim 5\times10^5~\text{m/s}$), $\phi=\frac{3\pi}{5}$, and take the Fermi energy to be in the range $\varepsilon_\text{F} = 5-20~\text{meV}$ so that $k_\text{F} \sim \varepsilon_\text{F}/\hbar v_\text{F} \sim 0.002-0.007~\text{\AA}^{-1}$. For simplicity, we restrict the calculation to the Fermi pocket of the upper band (a better Fermi surface reconstruction would require a UV complete model). Moreover, we set $\tau = 10^{-13}~\text{s}$ and assume $\omega$ is small enough that we can safely neglect $\omega \tau$. We diagonalize $H_P(\textbf{k})$ on a $100\times100\times100$ grid with $k_\text{max}=1.5k_\text{F}$ and compute the conductivity tensors found in App.~\ref{app:condFS}. 

\begin{figure}
    \centering
    \includegraphics[width=\linewidth]{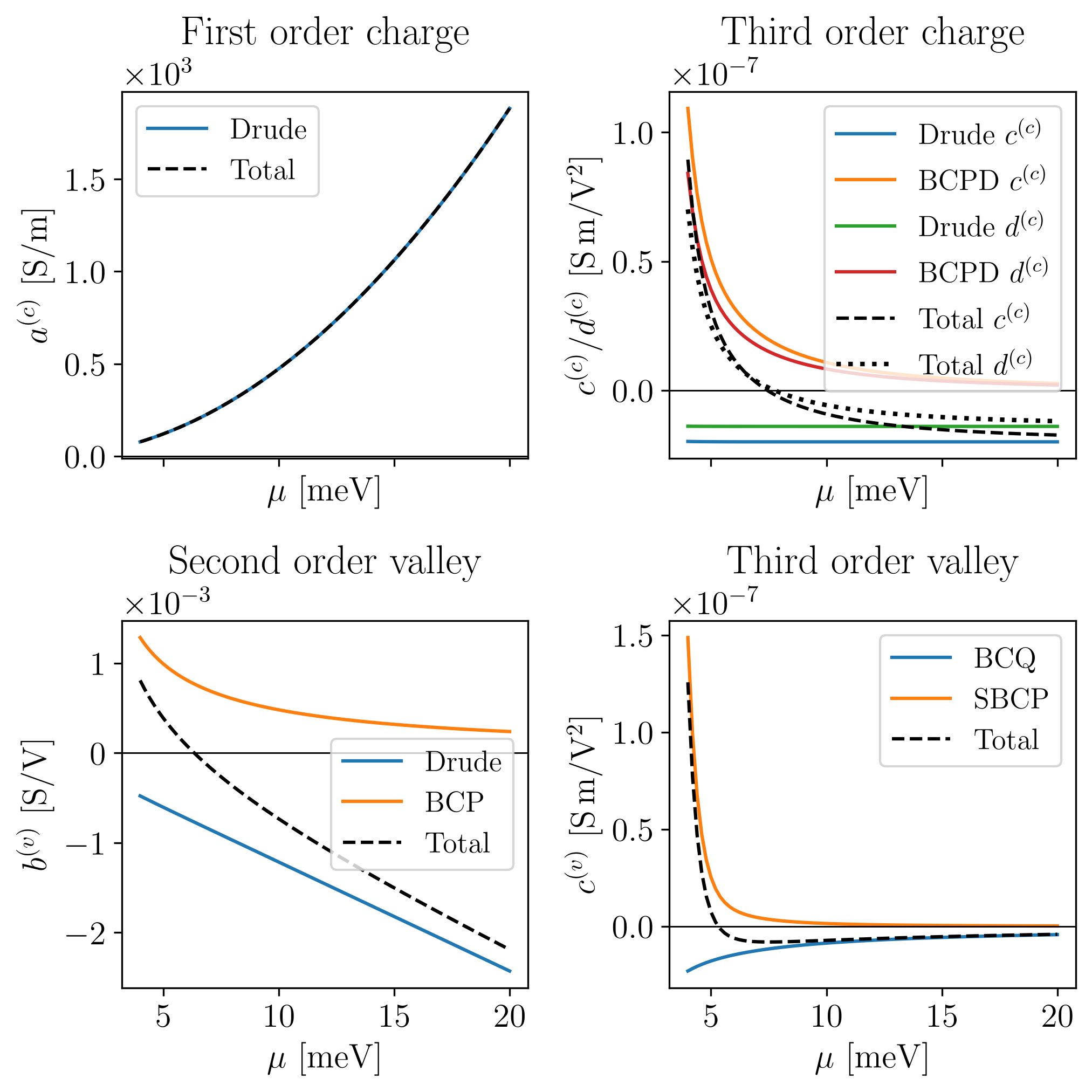}
    \caption{Non-zero charge and valley conductivity coefficients in three-fold crossings in space group $I4_132$, described by Eq.~\eqref{eq:kp199}, as a function of the chemical potential set above the crossing's energy. We set $\phi=\frac{3\pi}{5}$, $\hbar v_\text{F} = 3$ eV \AA, $k_BT = 10^{-4}$eV, and $\tau = 10^{-13}$s.}
    \label{fig:cvcurrents}
\end{figure}

Since the point group symmetry of the single-valley model is $T$, the most general form of the constitutive relations are:
\begin{equation}\label{Tconst}
\begin{cases}
    j_x^{(\eta)} = a^{(\eta)} E_x + b^{(\eta)} E_y E_z + (c_1^{(\eta)} E_y^2 + c_2^{(\eta)} E_z^2+c_3^{(\eta)} E_x^2)E_x \\ 
    j_y^{(\eta)} = a^{(\eta)} E_y + b^{(\eta)} E_x E_z + (c_1^{(\eta)} E_z^2 + c_2^{(\eta)} E_x^2 + c_3^{(\eta)} E_y^2)E_y \\ 
    j_z^{(\eta)} = a^{(\eta)} E_z + b^{(\eta)} E_y E_x + (c_1^{(\eta)} E_x^2 + c_2^{(\eta)} E_y^2 + c_3^{(\eta)} E_z^2)E_z
\end{cases},
\end{equation}
where $\eta = P, -P$ denotes the two valleys. One finds that under the action of the $C_{4z}$ rotation exchanging the valleys, $a^{(P)} = a^{(-P)}$, $b^{(P)} = -b^{(-P)}$, $c_3^{(P)} = c_3^{(-P)}$, and $c_1^{(P)} = c_2^{(-P)}$. We can then define the total charge and valley currents as $\textbf{j}^{(c)} = \textbf{j}^{(P)} + \textbf{j}^{(-P)}$ and $\textbf{j}^{(v)} = \textbf{j}^{(P)} - \textbf{j}^{(-P)}$ respectively, and similarly for $a^{(c),(v)} = a^{(P)} \pm a^{(-P)},b^{(c),(v)} = b^{(P)} \pm b^{(-P)},c^{(c),(v)} = c_1^{(P)} \pm c_1^{(-P)} = c_2^{(-P)} \pm c_2^{(P)}$ and $d^{(c),(v)} = c_3^{(P)} \pm c_3^{(-P)}$. Since the crystalline point group symmetry is $O$, the allowed charge and valley responses are thus:
\begin{equation}
\begin{cases}
    j_x^{(\lambda)} = a^{(\lambda)} E_x + b^{(\lambda)} E_y E_z+c^{(\lambda)} (E_y^2 \pm E_z^2)E_x + d^{(\lambda)} E_x^3 \\ 
    j_y^{(\lambda)} = a^{(\lambda)} E_y + b^{(\lambda)} E_x E_z+c^{(\lambda)} (E_z^2 \pm E_x^2)E_y + d^{(\lambda)} E_y^3 \\ 
    j_z^{(\lambda)} = a^{(\lambda)} E_z + b^{(\lambda)} E_y E_x+ c^{(\lambda)} (E_x^2 \pm E_y^2)E_z  + d^{(\lambda)} E_z^3 
\end{cases},
\end{equation}
where now $\lambda=c,v$ selects the type of response (even or odd under valley exchange), and $+$ is chosen for charge and $-$ for valley response. It follows that $a^{(v)} = b^{(c)} = d^{(v)}=0$. We plot the non-zero coefficients obtained numerically in Fig.~\ref{fig:cvcurrents}. 

Firstly, we find that the only surviving valley coefficients are $b^{(v)}$ and $c^{(v)}$, as expected from our previous symmetry arguments. Furthermore, we note that the second order valley current is generally not fully transverse, since $\textbf{E} \cdot \textbf{j}^{2,(v)} = 3b^{(v)} E_x E_y E_z$ which is only zero when the electric field is along special high symmetry directions. One must go to third order to obtain a fully transverse valley current, where $\textbf{E} \cdot \textbf{j}^{3,(v)}=0$. Secondly, because the contributions to the valley coefficients come in opposite signs and with different scalings in $\mu$, it is possible to reverse their sign by tuning the chemical potential. If we imagine applying an electric field to a thin slab made out of a multifold semimetal described by Eq.~\eqref{eq:kp199}, one could produce an accumulation of $P$ and $-P$ valley electrons on opposite faces of the device, as shown in Fig.~\ref{fig:coverfig}b. This in turn could result in an orbital magnetic moment gradient in the sample. Furthermore, the second- and third-order valley coefficients each change sign at some characteristic chemical potentials $\mu_{2,*}$ and $\mu_{3,*}$ determined by the competition between their microscopic contributions. Tuning the chemical potential above or below $\mu_{3,*}$, for example by gating the sample, one could then reverse the third-order valley current, thus acting as a gate-operated valley filter. 

\subsection{Chiral valve in space group $P2_13$}
We proceed with the analysis of multifold fermions in space group $P2_13$. At $\Gamma$ the crystalline symmetries can protect a four-fold crossing point that is described, to linear order, by a spin-$3/2$ fermion. When computed on small spherical Fermi surfaces around the node, the four bands carry Chern numbers $C_m=-2m$ with $m=\tfrac{3}{2},\tfrac{1}{2},-\tfrac{1}{2},-\tfrac{3}{2}$, i.e.\ $\{-3,-1,+1,+3\}$ which sum to zero. The topological charge of the node is therefore $|C_\Gamma|=|\sum_{\epsilon_m<0}C_m|=4$. At $R$, time-reversal symmetry combined with three-fold rotation protects a six-fold degeneracy that can be viewed as two spin-1 triplets related by time-reversal. Each triplet contributes $C_m=\{-2,0,+2\}$, giving a total node charge $|C_R|=4$. 

Our results directly apply to CoSi and PdGa in space group $P2_13$ that have been the subject of recent investigations. Complementary transport studies have considered non-linear thermoelectric responses in CoSi \cite{NakazawaEtAl2025}, non-linear and magnetotransport signatures of spin-3/2 Rarita–Schwinger–Weyl fermions \cite{MandalEtAl2025}, and the chiral-anomaly magnetoresponse of multifold fermions in CoSi \cite{BalduiniEtAl2024}. In particular, Ref.~\cite{Dixit2025} studies non-linear response in a three-legged device made from PdGa, which we briefly discuss.
An AC current at frequency $\omega$ was applied along the crystal $[100]$ axis, and a voltage at first, second and third harmonic frequencies $\omega, 2\omega, 3\omega$ were measured along the $[0\bar11]$ axis. To match the experimental set-up in Ref.~\cite{Dixit2025}, the current-field constitutive relation
\begin{equation}\label{currentfield}
    j_a= \sigma_{ab}^{(1)} E_b + \sigma_{abc}^{(2)} E_b E_c + \sigma_{abcd}^{(3)} E_b E_c E_d  + O(E^4)
\end{equation}
must be inverted into the form
\begin{equation}\label{fieldcurrent}
    E_a = \rho_{an}^{(1)} j_n + \rho_{anm}^{(2)} j_n j_m + \rho_{anml}^{(3)} j_n j_m j_l+O(j^4).
\end{equation}
In the case of the relevant point group $T$ we find that the electric field produced by applying a current can be decomposed as:
\begin{equation}
\begin{cases}
E_x= \alpha j_{x}+ \beta j_{y} j_{z}+\gamma_{1} j_{y}^2 j_{x} + \gamma_{2} j_{z}^2 j_{x} + \gamma_{3} j_{x}^3 \\[4pt] 
E_y= \alpha j_{y}+ \beta j_{x} j_{z}+\gamma_{1} j_{z}^2 j_{y} + \gamma_{2} j_{x}^2 j_{y} + \gamma_{3} j_{y}^3  \\[4pt] 
E_z = \alpha j_{z}+ \beta j_{x} j_{y}+\gamma_{1} j_{x}^2 j_{z} + \gamma_{2} j_{y}^2 j_{z} + \gamma_{3} j_{z}^3,
\end{cases}
\end{equation}
where the expression for $\alpha, \beta, \gamma_1, \gamma_2, \gamma_3$ can be found in App.~\ref{app:inverserel}. Thus, when a current is applied along $x$, only the electric fields $E_x = \alpha j_x + \gamma_3 j_x^3$ will result in a voltage. Thus, the experimentally observed transverse response in the $[0\bar11]$ direction is not found within our theory. It is noteworthy that the second and third harmonic transverse response observed in Ref.~\cite{Dixit2025} only appears after a threshold current is applied. This may indicate a current-induced symmetry-broken state in the device. Another explanation for the experimentally observed response is that it stems from effects not included in our derivation, such as impurity scattering or device geometry. The latter is most strongly breaking the point group symmetry at the device boundaries -- exactly where the arms in which the measured voltage builds up are attached. Surface states, which are not part of our bulk theory but abundant in topological semimetals, may also add to a boundary transport contribution.

\section{Conclusions}\label{sec:conclusion}

With precise device preparation, enabled through advanced growth methods for heterostructures and focused ion beam milling~\cite{Moll-FibWedge,Guo_switchable}, transport experiments are reaching a new level of control over the device geometry, opening the way for innovative measurement schemes of non-linear and tensorial response functions. For the interpretation of measurements, a careful study of the response functions in relation to the symmetries of crystal and the device geometry is required. We performed this analysis for the relevant case of multifold fermions in 3D time-reversal symmetric crystals, highlighting the opportunities for non-linear valley-tronics in these systems. For the three-fold crossings at the $P$ points in materials with space group $I4_132$, symmetry permits a second order valley response and enforces a purely transverse third order valley current whose sign can reverse as the chemical potential is varied. In a finite slab, this bulk current is expected to generate opposite surface valley chemical potentials.

The formalism we presented can be readily applied to other relevant fermiologies, such as magnetic multifold fermion semimetals~\cite{10.1063/1.5124314} where spin transport could additionally be investigated, or nodal line semimetals~\cite{Fang_2016}. It could furthermore be fully extended to the AC regime where topological contributions become relevant. Finite-field extensions may additionally connect to recent studies of the longitudinal magnetoconductance of pseudospin-1 fermions \cite{AhmadSharma2025}.

\section{Acknowledgements}\label{sec:acknow}
The authors thank Stuart S.~P.~Parkin and Anvesh Dixit for numerous interactions and for sharing their results ahead of publication in Ref.~\onlinecite{Dixit2025}. A.K.D. thanks Zoltán Guba and Wojciech J.~Jankowski for fruitful discussions. T.N. and A.K.D. acknowledge support from the Swiss National Science Foundation through a Consolidator Grant (iTQC, TMCG-2213805).
\bibliography{refs}

\begin{appendix}
\section{AC distribution functions}\label{app:ACdistr}

Let us take a monochromatic external AC field $E_\alpha=\Re[\mathcal{E}_\alpha e^{i\omega t}]$, and expand $f_n$ into harmonics as $f_n = \sum_m f_{n,m} e^{im\omega t}$ with $f_{n,m}^*=f_{n,-m}$ to ensure that $f_n$ is real-valued. Inserting this into Eq.~\eqref{eq:frecursive}, the Fourier-transformed modes can then be obtained recursively from
\begin{align}
 f_{n,m} = \frac{e\tau}{2\hbar}\frac{1}{1+im\omega \tau}(\mathcal{E}_\alpha \partial_\alpha f_{n-1,m-1}+\mathcal{E}_\alpha^* \partial_\alpha f_{n-1,m+1}).
\end{align}
The latter equation, together with the condition that $f_{0,m}=f_0 \delta_{m,0}$ shows that $f_{n,m}$ will be non-zero for $m=n,n-2,n-4,...$. For reference, we provide the explicit form of $f_1, f_2$ and $f_3$ (which match with what was obtained in Ref.~\cite{zhang2023higher}). The first order correction is directly found to be:
\begin{equation}\label{f1}
  f_{1,m}= \frac{e \tau}{2\hbar}\bigg(\frac{\mathcal{E}_\alpha\partial_\alpha f_0}{1+i\omega \tau} \delta_{m,1} + \frac{\mathcal{E}^*_\alpha\partial_\alpha f_0}{1-i\omega \tau} \delta_{m,-1}\bigg),
\end{equation}
which gives rise to the familiar Drude conductivity formula for the first harmonic current generation. The second order corrections contain zeroth and second harmonics $f_{2,m}=f_{2,0}\delta_{m,0}+f_{2,2}\delta_{m,2} + f^*_{2,2}\delta_{m,-2}$ where
\begin{subequations}
\begin{align}
  &f_{2,0} = \frac{(e\tau/2\hbar)^2}{(1+i\omega \tau)}\mathcal{E}_\alpha^*\mathcal{E}_\beta \partial_\alpha \partial_\beta f_0 + c.c, \label{f2_1}\\ 
  &f_{2,2}= \frac{(e\tau/2\hbar)^2}{(1+i\omega \tau)(1+2i\omega \tau)}\mathcal{E}_\alpha\mathcal{E}_\beta \partial_\alpha \partial_\beta f_0.\label{f2_2}
\end{align}
\end{subequations}
Finally the third order correction is $f_{3,m}=f_{3,1}\delta_{m,1}+f_{3,3}\delta_{m,3} +f^*_{3,1}\delta_{m,-1} +f^*_{3,3}\delta_{m,-3}$ where
\begin{subequations}
\begin{align}
    &f_{3,3} =  \frac{(e\tau/2\hbar)^3}{(1+i\omega \tau)(1+2i\omega \tau)(1+3i\omega \tau)}\mathcal{E}_\alpha \mathcal{E}_\beta \mathcal{E}_\gamma \partial_\alpha \partial_\beta \partial_\gamma f_0, \label{f3_1}\\
    &f_{3,1} = \frac{3(e\tau/2\hbar)^3}{(1-i\omega \tau)(1+i\omega \tau)(1+2i\omega \tau)}\mathcal{E}_\alpha^* \mathcal{E}_\beta \mathcal{E}_\gamma \partial_\alpha \partial_\beta \partial_\gamma f_0.\label{f3_2}
\end{align}
\end{subequations}
\section{Proof of Eq.~\eqref{HFeq}}\label{app:HF}
Suppose the crystal in the absence of an external electric field is described by the Hamiltonian $H_0$. We can turn on the electric field adiabatically by introducing the perturbation $\lambda e\textbf{E}\cdot \textbf{r}$ where $\lambda \in [0,1]$. We let $\ket{\Psi(\lambda)}$ be the field-dressed state adiabatically connected to the active-band state at $\lambda=0$. The perturbative energy is then
\begin{equation}\label{energy1}
    \varepsilon(\lambda) = \braket{\Psi(\lambda)|H(\lambda)|\Psi(\lambda)}=\varepsilon^{(0)} + \lambda e \textbf{E} \cdot \textbf{r}_c^{(0)} + \sum_{n=2}^{\infty} \lambda^n \varepsilon^{(n)},
\end{equation}
where we defined the wavepacket center using the positional shifts:
\begin{equation}
    \textbf{r}_c(\lambda) = \braket{\Psi(\lambda)|\textbf{r}|\Psi(\lambda)} = \textbf{r}_c^{(0)} + \sum_{m=0}^\infty \lambda^m \boldsymbol{\mathcal{A}}^{(m)}.
\end{equation}
Using the Feynman-Hellmann theorem, we find that
\begin{equation}
    \frac{d\varepsilon(\lambda)}{d\lambda} =e\textbf{E}\cdot \textbf{r}_c = e\textbf{E}\cdot \textbf{r}_c^{(0)} +  \sum_{m=1}^\infty \lambda^m e\textbf{E}\cdot  \boldsymbol{\mathcal{A}}^{(m)}.
\end{equation}
Meanwhile, if one differentiates Eq.~\eqref{energy1} then one gets:
\begin{equation}
    \frac{d\varepsilon(\lambda)}{d\lambda} = e\textbf{E}\cdot \textbf{r}_c^{(0)} + \sum_{n=2}^\infty n\lambda^{n-1}\varepsilon^{(n)}.
\end{equation}
Comparing term by term results in the desired identity:
\begin{equation}
    \varepsilon^{(n)} = \frac{1}{n} e\textbf{E}\cdot \boldsymbol{\mathcal{A}}^{(n-1)}.
\end{equation}
An alternative, even shorter proof goes as follows. Consider the Hamiltonian $H=H_0+e\textbf{E}\cdot \textbf{r}$. We again have from Feynman-Hellmann that
\begin{equation}
    \frac{\partial \varepsilon}{\partial E_\alpha} = e \braket{r_\alpha} \implies \frac{\partial \varepsilon^{(n)}}{\partial E_\alpha} = e A_\alpha^{(n-1)}.
\end{equation}
Moreover, since $\varepsilon^{(n)}$ is homogeneous of degree $n$ in $E_\alpha$, we have from Euler's theorem that
\begin{equation}
    E_\alpha \frac{\partial \varepsilon^{(n)}}{\partial E_\alpha} = n \varepsilon^{(n)},
\end{equation}
from which one can again retrieve Eq.~\eqref{HFeq}. 
\section{Inverse constitutive relations}\label{app:inverserel}
We start with the general current-field relation
\begin{equation}\label{currentfield}
    j_a= \sigma_{ab}^{(1)} E_b + \sigma_{abc}^{(2)} E_b E_c + \sigma_{abcd}^{(3)} E_b E_c E_d  + O(E^4),
\end{equation}
which the point group $T$ reduces to:
\begin{equation}
\begin{cases}
j_x= a E_{x}+b E_{y} E_{z}+c_{1} E_{y}^2 E_{x} + c_{2} E_{z}^2 E_{x} + c_{3} E_{x}^3, \\[4pt] 
j_y= a E_{y}+b E_{x} E_{z}+c_{1} E_{z}^2 E_{y} + c_{2} E_{x}^2 E_{y} + c_{3} E_{y}^3,  \\[4pt] 
j_z = a E_{z}+b E_{x} E_{y}+c_{1} E_{x}^2 E_{z} + c_{2} E_{y}^2 E_{z} + c_{3} E_{z}^3.
\end{cases}
\end{equation}
We wish to invert these relations into the field-current form
\begin{equation}\label{fieldcurrent}
    E_a = \rho_{an}^{(1)} j_n + \rho_{anm}^{(2)} j_n j_m + \rho_{anml}^{(3)} j_n j_m j_l+O(j^4).
\end{equation}
Substituting Eq.~\eqref{fieldcurrent} into Eq.~\eqref{currentfield} we get
\begin{equation}
    \begin{cases}
        \sigma_{ab}^{(1)} \rho_{bn}^{(1)}=\delta_{na},\\[4pt]
        \sigma_{ab}^{(1)}\rho_{bnm}^{(2)} +\sigma_{abc}^{(2)} \rho_{bn}^{(1)} \rho_{cm}^{(1)} =0,\\[4pt]
        \sigma_{ab}^{(1)} \rho_{bnml}^{(3)} + \sigma_{abcd}^{(3)} \rho_{bn}^{(1)} \rho_{cm}^{(1)} \rho_{dl}^{(1)} + \sigma_{abc}^{(2)}(\rho_{bn}^{(1)} \rho_{cml}^{(2)} + \rho_{bnm}^{(2)} \rho_{cl}^{(1)}) = 0.
    \end{cases} 
\end{equation}
Assuming that $\sigma^{(1)}$ is invertible, we find that $\rho^{(1)} = (\sigma^{(1)})^{-1}$, and thus
\begin{align}
    &\rho^{(2)}_{anm} =-\rho_{ai}^{(1)} \sigma^{(2)}_{ibc} \rho_{bn}^{(1)} \rho_{cm}^{(1)},\nonumber\\
    &\rho^{(3)}_{anml} = -\rho_{ai}^{(1)}\sigma^{(3)}_{ibcd} \rho_{bn}^{(1)} \rho_{cm}^{(1)} \rho_{dl}^{(1)} - \rho_{ai}^{(1)}\sigma_{ibc}^{(2)}(\rho_{bn}^{(1)} \rho_{cml}^{(2)} + \rho_{bnm}^{(2)} \rho_{cl}^{(1)}).
\end{align}
In the case of the relevant point group $T$ we obtain
\begin{equation}
\begin{cases}
E_x= \alpha j_{x}+ \beta j_{y} j_{z}+\gamma_{1} j_{y}^2 j_{x} + \gamma_{2} j_{z}^2 j_{x} + \gamma_{3} j_{x}^3, \\[4pt] 
E_y= \alpha j_{y}+ \beta j_{x} j_{z}+\gamma_{1} j_{z}^2 j_{y} + \gamma_{2} j_{x}^2 j_{y} + \gamma_{3} j_{y}^3,  \\[4pt] 
E_z = \alpha j_{z}+ \beta j_{x} j_{y}+\gamma_{1} j_{x}^2 j_{z} + \gamma_{2} j_{y}^2 j_{z} + \gamma_{3} j_{z}^3,
\end{cases}
\end{equation}
where
\begin{subequations}\label{invconst}
\begin{align}
    &\alpha = \frac{1}{a}, \ \beta = -\frac{b}{a^3},\label{invconst_1}\\
    &\gamma_1=\frac{b^2}{a^5}-\frac{c_1}{a^4}, \ \gamma_2 = \frac{b^2}{a^5}-\frac{c_2}{a^4}, \ \gamma_3 = -\frac{c_3}{a^4}.\label{invconst_2}
\end{align}
\end{subequations}
\section{Conductivities as Fermi surface quantities}\label{app:condFS}
For convenience, we collect the expressions for the various zero-temperature conductivities. For each type of contribution, we provide a Fermi surface integral formula involving $\int_\text{FS} \equiv \frac{1}{(2\pi)^3}\int_\text{FS}dS_k$. If there are multiple active bands, one must sum over all the disjoint Fermi surface integrals. The two contributions to the linear conductivity are:
\begin{align}
    \sigma_{\alpha \beta}^{(1),\text{D}} &= \frac{e}{\hbar}  \bigg(\frac{e\tau}{\hbar}\bigg) \int_\textbf{k} \partial_\alpha \varepsilon^{(0)} \partial_\beta \varepsilon^{(0)} \delta(\varepsilon^{(0)}-\varepsilon_\text{F}) \nonumber\\
    &=\frac{e}{\hbar} \bigg(\frac{e\tau}{\hbar}\bigg) \oint_\text{FS} \partial_\alpha \varepsilon^{(0)} \frac{\partial_\beta \varepsilon^{(0)}}{|\nabla \varepsilon^{(0)}|}, 
\end{align}
and
\begin{align}
    \sigma_{\alpha \beta}^{(1),\text{BC}} &= -\frac{e^2}{\hbar} \int_\textbf{k} (A^{(0)}_\beta \partial_\alpha\varepsilon^{(0)}-A^{(0)}_\alpha\partial_\beta\varepsilon^{(0)})\delta(\varepsilon^{(0)}-\varepsilon_\text{F}) \nonumber\\
    &=-\frac{e^2}{\hbar} \oint_\text{FS} \bigg(A^{(0)}_\beta \frac{\partial_\alpha \varepsilon^{(0)}}{|\nabla \varepsilon^{(0)}|} -A^{(0)}_\alpha\frac{\partial_\beta \varepsilon^{(0)}}{|\nabla \varepsilon^{(0)}|}\bigg).
\end{align}
Note the integrand of the latter is only gauge-invariant up to a total derivative. Moreover, since $A_\alpha^{(0)}$ cannot be defined smoothly on a Fermi surface with non-trivial Chern number, one must express $\mathcal{A}$ on separate patches, and take into account contributions from the patch boundaries. The contributions to the second order conductivity are:
\begin{align}
    \sigma_{\alpha \beta \gamma}^{(2),\text{D}} &= -\frac{e}{\hbar} \bigg(\frac{e\tau}{\hbar}\bigg)^2 \int_\textbf{k} \partial_\alpha \partial_\beta \varepsilon^{(0)} \partial_\gamma \varepsilon^{(0)} \delta(\varepsilon^{(0)}-\varepsilon_\text{F}) \nonumber \\
    &=-\frac{e}{\hbar}\bigg(\frac{e\tau}{\hbar}\bigg)^2   \oint_\text{FS} \partial_\alpha \partial_\beta \varepsilon^{(0)} \frac{\partial_\gamma\varepsilon^{(0)}}{|\nabla \varepsilon^{(0)}|},
\end{align}
and 
\begin{align}
    \sigma_{\alpha \beta \gamma}^{(2),\text{BCD}} &= \frac{e^2}{\hbar} \bigg(\frac{e\tau}{\hbar}\bigg) \int_\textbf{k} F_{\alpha\gamma}^{(0)} \partial_\beta \varepsilon^{(0)} \delta(\varepsilon^{(0)}-\varepsilon_\text{F}) \nonumber \\
    &=\frac{e^2}{\hbar}\bigg(\frac{e\tau}{\hbar}\bigg)   \oint_\text{FS} F_{\alpha\gamma}^{(0)} \frac{\partial_\beta\varepsilon^{(0)}}{|\nabla \varepsilon^{(0)}|},
\end{align}
(where we defined $F^{(0)}_{\alpha \beta} = \partial_\alpha A^{(0)}_\beta - \partial_\beta A^{(0)}_\alpha$) and
\begin{align}
    \sigma_{\alpha \beta \gamma}^{(2),\text{BCP}} &= -\frac{e^3}{2\hbar}  \int_\textbf{k} (\mathcal{G}^{(1)}_{\beta \gamma}\partial_\alpha \varepsilon^{(0)}-\mathcal{G}^{(1)}_{\alpha \gamma}\partial_\beta\varepsilon^{(0)}-\mathcal{G}^{(1)}_{\alpha \beta}\partial_\gamma\varepsilon^{(0)}) \nonumber\\
    &\hspace{2cm}\times \delta(\varepsilon^{(0)}-\varepsilon_\text{F}) \nonumber \\
    &=-\frac{e^3}{2\hbar}  \oint_\text{FS} \bigg(\mathcal{G}^{(1)}_{\beta \gamma}\frac{\partial_\alpha\varepsilon^{(0)}}{|\nabla\varepsilon^{(0)}|} -\mathcal{G}^{(1)}_{\alpha \gamma}\frac{\partial_\beta\varepsilon^{(0)}}{|\nabla\varepsilon^{(0)}|} -\mathcal{G}^{(1)}_{\alpha \beta}\frac{\partial_\gamma\varepsilon^{(0)}}{|\nabla\varepsilon^{(0)}|} \bigg).
\end{align}
Finally, the contributions to the third order conductivity are:
\begin{align}
    \sigma_{\alpha \beta \gamma \eta}^{(3),\text{D}} &= \frac{e}{\hbar} \bigg(\frac{e\tau}{\hbar}\bigg)^3 \int_\textbf{k} \partial_\alpha \partial_\beta \partial_\gamma \varepsilon^{(0)} \partial_\eta \varepsilon^{(0)} \delta(\varepsilon^{(0)}-\varepsilon_\text{F}) \nonumber \\
    &=\frac{e}{\hbar}\bigg(\frac{e\tau}{\hbar}\bigg)^3   \oint_\text{FS} \partial_\alpha \partial_\beta \partial_\gamma \varepsilon^{(0)} \frac{\partial_\eta\varepsilon^{(0)}}{|\nabla \varepsilon^{(0)}|},
\end{align}
and 
\begin{align}
    \sigma_{\alpha \beta \gamma \eta}^{(3),\text{BCQ}} &= -\frac{e^2}{\hbar} \bigg(\frac{e\tau}{\hbar}\bigg)^2\int_\textbf{k} \partial_\beta F_{\alpha\gamma}^{(0)} \partial_\eta \varepsilon^{(0)}  \delta(\varepsilon^{(0)}-\varepsilon_\text{F}) \nonumber \\
    &=-\frac{e^2}{\hbar}\bigg(\frac{e\tau}{\hbar}\bigg)^2   \oint_\text{FS} \partial_\beta F_{\alpha\gamma}^{(0)} \frac{\partial_\eta\varepsilon^{(0)}}{|\nabla \varepsilon^{(0)}|},
\end{align}
and
\begin{align}
    \sigma_{\alpha \beta \gamma \eta}^{(3),\text{BCPD}} &= \frac{e^3}{2\hbar}   \bigg(\frac{e\tau}{\hbar}\bigg)\int_\textbf{k} (\partial_\alpha \mathcal{G}^{(1)}_{\beta \gamma}-\partial_\beta \mathcal{G}^{(1)}_{\alpha \gamma}-\partial_\gamma\mathcal{G}^{(1)}_{\alpha \beta})\nonumber\\
    &\hspace{3cm}\times\partial_\eta\varepsilon^{(0)} 
 \delta(\varepsilon^{(0)}-\varepsilon_\text{F}) \nonumber \\
    &=\frac{e^3}{2\hbar}  \bigg(\frac{e\tau}{\hbar}\bigg) \oint_\text{FS} (\partial_\alpha \mathcal{G}^{(1)}_{\beta \gamma}-\partial_\beta \mathcal{G}^{(1)}_{\alpha \gamma}-\partial_\gamma\mathcal{G}^{(1)}_{\alpha \beta})\nonumber\\
    &\hspace{3cm}\times\frac{\partial_\eta\varepsilon^{(0)}}{|\nabla \varepsilon^{(0)}|},
\end{align}
and 
\begin{align}
    \sigma_{\alpha \beta \gamma \eta}^{(3),\text{SBCP}} &= -\frac{e^4}{3\hbar}  \int_\textbf{k} (\mathcal{G}^{(2)}_{\beta \gamma \eta}\partial_\alpha \varepsilon^{(0)}-\mathcal{G}^{(2)}_{\alpha \gamma \eta}\partial_\beta\varepsilon^{(0)}-\mathcal{G}^{(2)}_{\alpha \beta \eta}\partial_\gamma\varepsilon^{(0)} \nonumber\\
    &\hspace{2cm}-\mathcal{G}^{(2)}_{\alpha \beta \gamma}\partial_\eta\varepsilon^{(0)})\delta(\varepsilon^{(0)}-\varepsilon_\text{F}) \nonumber \\
    &=-\frac{e^4}{3\hbar}  \oint_\text{FS} \bigg(\mathcal{G}^{(2)}_{\beta \gamma \eta}\frac{\partial_\alpha\varepsilon^{(0)}}{|\nabla\varepsilon^{(0)}|} -\mathcal{G}^{(2)}_{\alpha \gamma \eta}\frac{\partial_\beta\varepsilon^{(0)}}{|\nabla\varepsilon^{(0)}|} \nonumber\\ 
    &\hspace{1.8cm}-\mathcal{G}^{(2)}_{\alpha \beta \eta}\frac{\partial_\gamma\varepsilon^{(0)}}{|\nabla\varepsilon^{(0)}|}-\mathcal{G}^{(2)}_{\alpha \beta \gamma}\frac{\partial_\eta\varepsilon^{(0)}}{|\nabla\varepsilon^{(0)}|} \bigg).
\end{align}
\section{Idealized multifold fermions}\label{app:multifold_kS}
To discuss multifold chiral fermions in full generality, one can consider the idealized isotropic limit of a multifold crossing hosting spin-$S$ fermions by taking the following linearized low-energy Hamiltonian:
\begin{equation}
    H(\textbf{k}) = \hbar v_\text{F} \textbf{k} \cdot \textbf{S}.
\end{equation}
This Hamiltonian possesses SO(3) symmetry where the Hamiltonian transforms under the spin-$S$ SU(2) representation:
\begin{equation}
    U_R H(\textbf{k}) U_R^\dagger = H(R\textbf{k}), \ \forall R \in \text{SO(3)}, \ U_R = e^{-i\mathbf{\theta}\cdot \textbf{S}},
\end{equation}
where $\mathbf{\theta}$ parametrizes the rotation $R$. This model can thus be viewed as the idealized limit of an isotropic multifold crossing. The energy bands are $\varepsilon_m^{(0)}=m\hbar v_\text{F} k$ where $m=-S,-S+1,...,S$ are the band indices and $k=|\textbf{k}|$. We thus find for the $m$th band that:
\begin{align}
    &\partial_i \varepsilon^{(0)} =m \hbar v_\text{F} n_i, \ \partial_i \partial_j \varepsilon = \frac{m\hbar v_\text{F}}{k} P_{ij}, \\
    &\partial_i \partial_j \partial_l \varepsilon^{(0)} = \frac{m\hbar v_\text{F}}{k^2}(3n_in_jn_l - \delta_{ij}n_l - \delta_{il}n_j - \delta_{jl} n_i),
\end{align}
where we defined $n_i = \frac{k_i}{k}$ and $P_{ij} = k\partial_i n_j = \delta_{ij}-n_in_j$. Next we compute the quantum geometric tensor:
\begin{align}
    Q_{ij} &= \bra{\partial_i u_m}(1-\ket{u_m}\bra{u_m})\ket{\partial_ju_m}\\
    &= \sum_{n\neq m} \braket{\partial_i u_m|u_n}\braket{u_n|\partial_j u_m}\\
    &= \sum_{n\neq m} \frac{\braket{u_m|\partial_i H|u_n}\braket{u_n|\partial_j H|u_m}}{(\epsilon_n-\epsilon_m)^2}\\
    &= \sum_{n\neq m} \frac{\braket{u_m|S_i|u_n}\braket{u_n|S_j|u_m}}{k^2(n-m)^2}.
\end{align}
Picking the quantization axis $\textbf{k}=k\hat{\textbf{z}}$ then the eigenstates are simultaneous eigenstates of both $S^2$ and $S_z$ i.e.~$\ket{u_m} = \ket{S,m}$. Consequently one finds that:
\begin{equation}
    Q(k\hat{\textbf{z}}) = \frac{1}{2k^2}\begin{pmatrix}
        S(S+1)-m^2 & im & 0 \\
        -im & S(S+1)-m^2 & 0 \\
        0 & 0 & 0
    \end{pmatrix},
\end{equation}
using the action of $S^+$ and $S^-$ on $\ket{S,m}$. The tensor at arbitrary $\textbf{k}$ is obtained by exploiting the SO(3) symmetry and rotating the tensor: $Q(\textbf{k}) = R_\textbf{n} Q(k\hat{\textbf{z}}) R_\textbf{n}^T$. Consequently we find that 
\begin{align}
    Q_{ij}(\textbf{k}) &= R_{ia} R_{jb} Q_{ab}(k\hat{\textbf{z}})\\
    &= \frac{S(S+1)-m^2}{2k^2}(R_{ix}R_{jx}+R_{iy}R_{jy}) \nonumber \\
    &\quad+ \frac{im}{2k^2}(R_{ix}R_{jy} - R_{iy} R_{jx})\\
    &= \frac{S(S+1)-m^2}{2k^2}P_{ij} + \frac{im}{2k^2}\epsilon_{ijl}n_l,
\end{align}
where we used the identities $R_{ix}R_{jx}+R_{iy}R_{jy} = P_{ij}$ and $R_{ix}R_{jy} - R_{iy} R_{jx} = \epsilon_{ijl}n_l$ for a rotation obeying $R\hat{\textbf{z}} = \hat{\textbf{n}}$. We can then read off the quantum metric and Berry curvature as
\begin{align}
    &g_{ij} = \frac{S(S+1)-m^2}{2k^2}P_ij, \\
    &F_{ij} = -\frac{m}{k^2}\epsilon_{ijl}n_l \implies \Omega_i = -\frac{m}{k^2}n_i,
\end{align}
respectively, with dipoles given by
\begin{align}
    &\partial_l g_{ij} = -\frac{S(S+1)-m^2}{2k^3}(2n_l P_{ij} + n_i P_{lj} + n_j P_{li}), \\ &\partial_l F_{ij} = \frac{m}{k^3}(3n_l\epsilon_{ijk}n_k - \epsilon_{lij}) \implies \partial_i \Omega_j = \frac{m}{k^3}(3n_i n_j - \delta_{ij}).
\end{align}
It is thus apparent that a Fermi surface from the $m$th band carries Chern number $C_m = -2m$. Indeed when the chemical potential lies near the multifold fermion degeneracies, small electron- or hole-like Fermi pockets form; each pocket encloses a multifold crossing point, and the Berry flux through the corresponding Fermi surface is quantized to give the topological charge of the surface $\oint_{S_\text{F}}\!\boldsymbol{\Omega}_m\cdot d\mathbf{S}=2\pi C_m$. The topological charge of the node is then given by the sum of the Fermi surface Chern numbers $C_\text{node} =  \sum_{m\text{ occ}} C_m$. The first BCP tensor can also be computed analogously:
\begin{equation}
    \mathcal{G}_{ij}^{(1)} = 2\text{Re}\sum_{n \neq m} \frac{\braket{u_m|\partial_i H|u_n}\braket{u_n|\partial_jH|u_m}}{(\varepsilon_m - \varepsilon_n)^3} = \frac{m}{\hbar v_\text{F} k^3} P_{ij}.
\end{equation}
Its dipole is given by:
\begin{equation}
    \partial_l  \mathcal{G}_{ij}^{(1)} = \frac{m}{\hbar v_\text{F} k^4}(5n_ln_in_j-3n_l\delta_{ij}-\delta_{li}n_j - \delta_{lj}n_i).
\end{equation}
Finally after some lengthy algebra one also finds that the second BCP tensor is
\begin{equation}
    \mathcal{G}_{ijl}^{(2)} = \frac{m}{2\hbar^2 v_\text{F}^2 k^5}n_i \epsilon_{jla} n_a,
\end{equation}
which is antisymmetric in $j,l$. Therefore the second order electric field induced positional shift vanishes $\mathcal{A}_\alpha^{(2)}=0$, as do the corresponding contributions to the conductivity.

Inserting these quantities into the expressions for the conductivities and integrating one readily obtains the expressions in Eq.~\eqref{eq:kscond1}--\eqref{eq:kscond3}. 
\section{Fermi surface distribution of conductivities}
We set $\hbar v_\text{F} = 3~\text{eV \AA}$ (corresponding to $v_\text{F} \sim 5\times10^5~\text{m/s}$), $\phi=\frac{3\pi}{5}$, and set the Fermi energy to $\varepsilon_\text{F} = 10 ~\text{meV}$ so that $k_\text{F} \sim \varepsilon_\text{F}/\hbar v_\text{F} \sim 0.003~\text{\AA}^{-1}$. For simplicity, we restrict the calculation to the Fermi pocket of the upper band (a better Fermi surface reconstruction would require a UV complete model). Moreover, we set $\tau = 10^{-13}~\text{s}$ and assume $\omega$ is small enough that we can safely neglect $\omega \tau$. We diagonalize $H_P(\textbf{k})$ on a $100\times100\times100$ grid with $k_\text{max}=1.5k_\text{F}$ and compute the conductivities found in App.~\ref{app:condFS}. We plot the contribution to the conductivity Fermi surface integrals at each \textbf{k} point belonging to the Fermi surface.
\subsubsection{First order response}
There are two separate contributions to the first order conductivity, a Drude conductivity $\sigma^{(1),\text{D}}_{\alpha\beta}$ and an intrinsic anomalous Hall conductivity due to the Berry curvature of the occupied states, $\sigma^{(1),\text{BC}}_{\alpha \beta}$. However, the point group symmetry forces the latter contribution to sum to zero as it cannot contribute to $a$ in Eq.~\eqref{Tconst}, the only allowed linear conductivity matrix element. Both of these quantities can be expressed as Fermi surface integrals (see App.~\ref{app:condFS}), but in the latter case the integrand is gauge-variant and does not carry physical meaning. In Fig.~\ref{fig:firstsecond order} we plot the distribution of the contributions to the conductivity over the Fermi pocket, so that when summed over the Fermi pocket, the result is the total Drude conductivity $\sigma^{(1),\text{D}}_{xx}$. 
\subsubsection{Second order response}
\begin{figure*}[h!] 
    \centering 
    \includegraphics[width=0.49\linewidth]{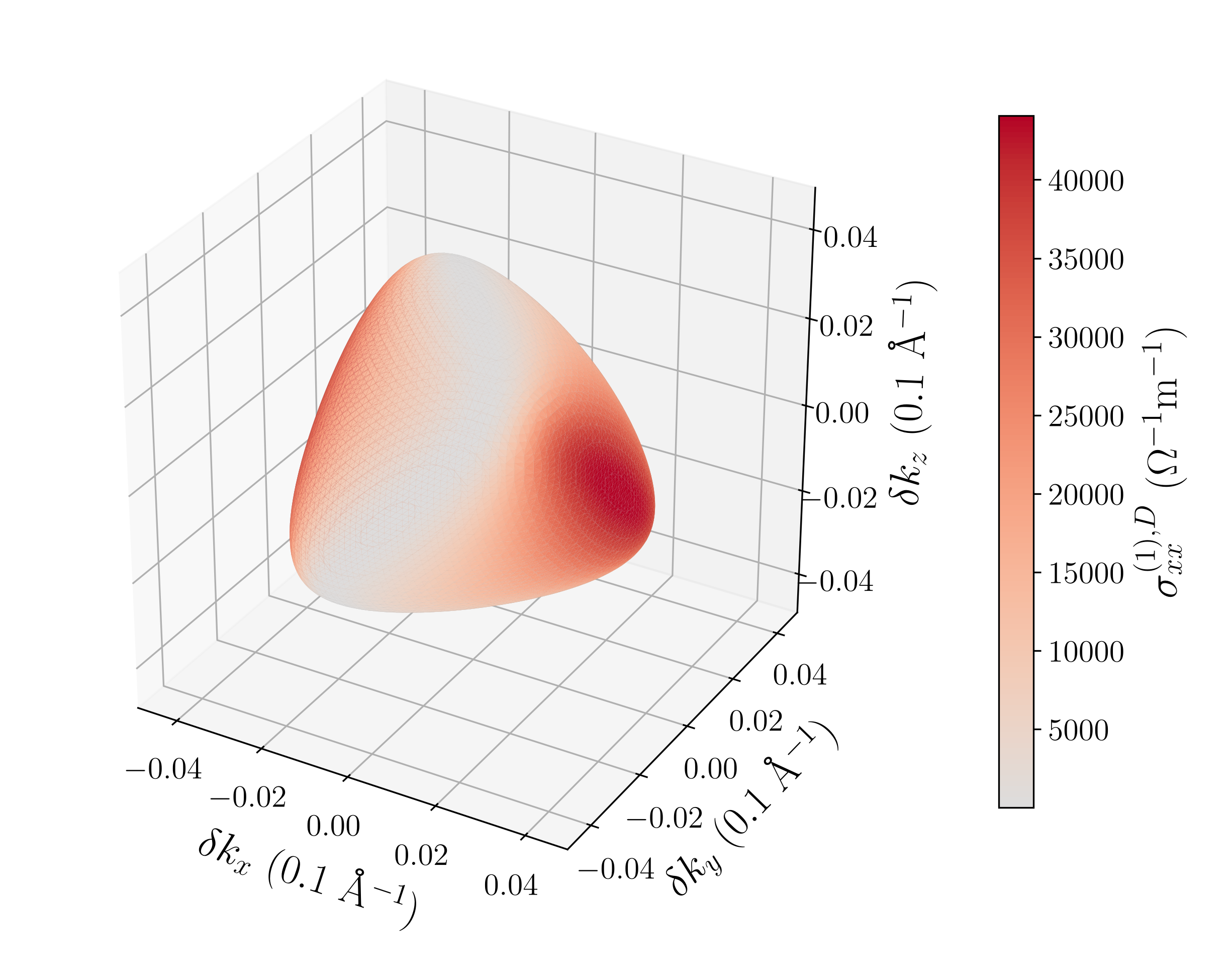}
    \includegraphics[width=0.49\linewidth]{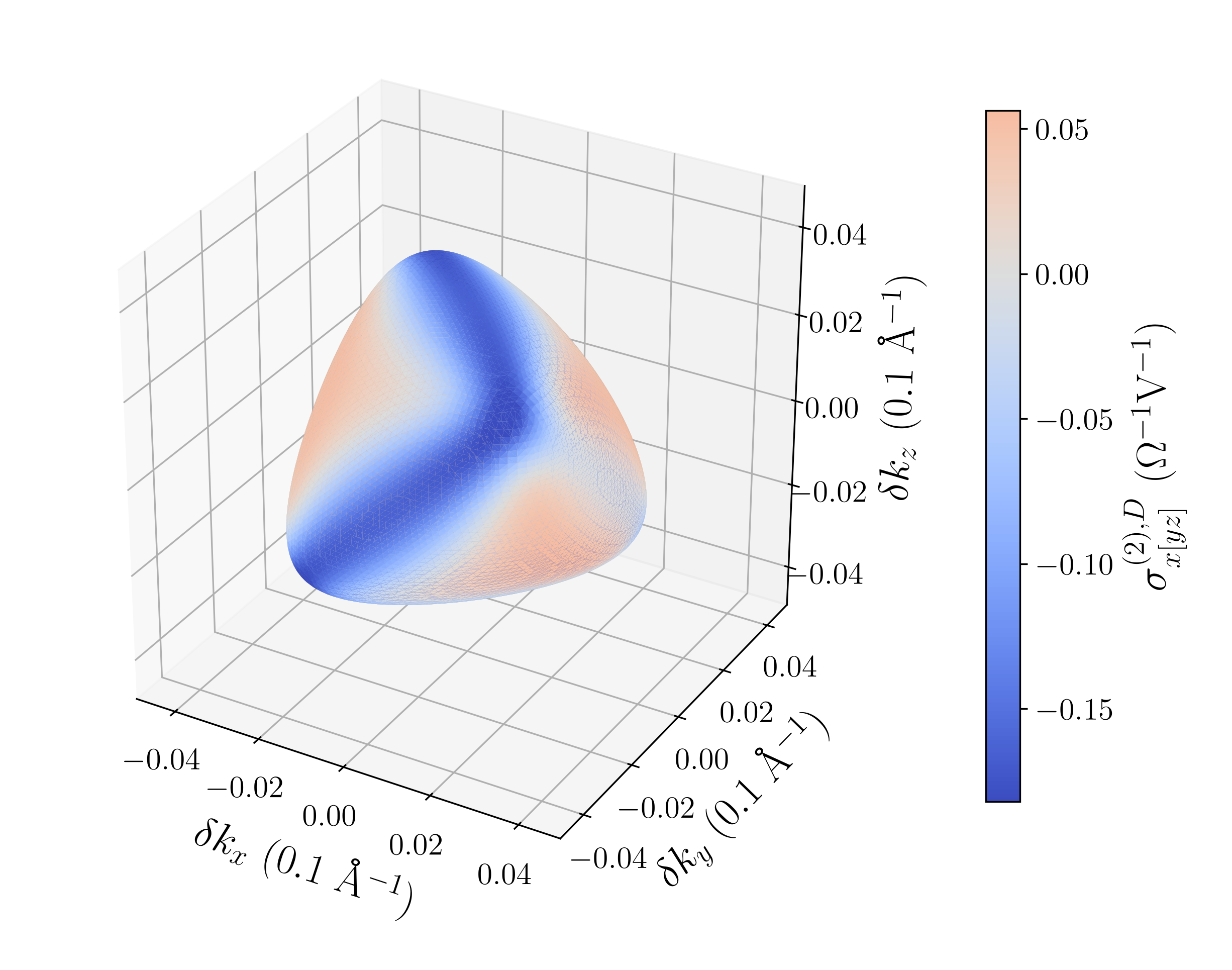} 
    \includegraphics[width=0.49\linewidth]{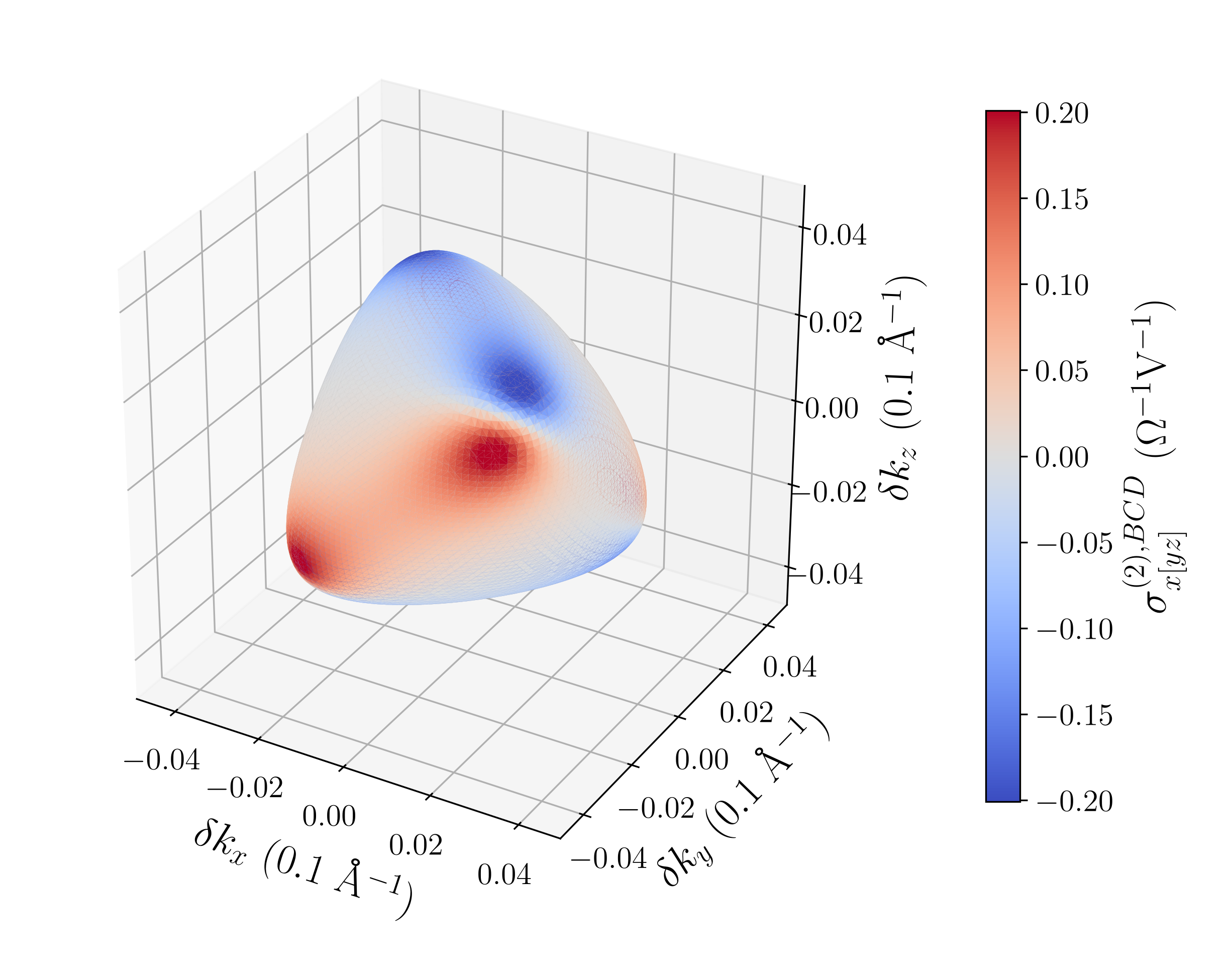} 
    \includegraphics[width=0.49\linewidth]{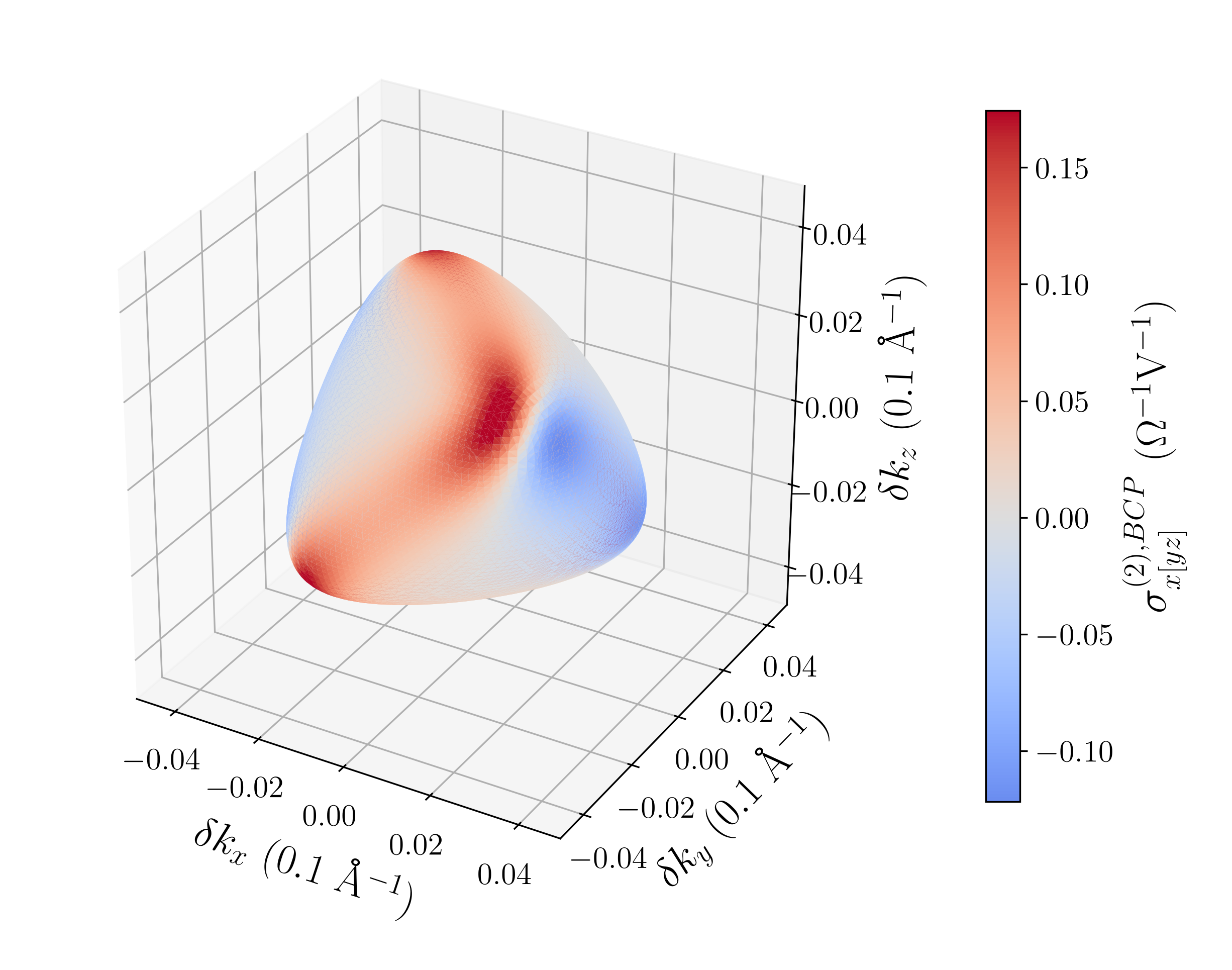} 
    \caption{$\textbf{k}$-resolved contributions to the first and second order conductivity on the Fermi pocket centered at $P$, coming from the linear Drude conductivity  $\sigma^{(1),\text{D}}_{xx}$, the second order Drude conductivity $\sigma^{(2),\text{D}}_{\alpha \beta \gamma}$, the Berry curvature dipole conductivity $\sigma^{(2),\text{BCD}}_{\alpha \beta \gamma}$ and the intrinsic berry connection polarizability conductivity $\sigma^{(2),\text{BCP}}_{\alpha \beta \gamma}$. When summed over the entire Fermi surface, these give $a_i$ and $b_1$ in Eq.~\eqref{jEgeneral1}. The only contributions which can sum to a finite value are $\sigma^{(1)}_{aa}$ and $\sigma^{(2)}_{a[bc]}$ where $abc$ are a permutation of $xyz$, and where $[...]$ in the subscript denotes index symmetrization (i.e. $\sigma^{(2)}_{a[bc]} = \sigma^{(2)}_{abc}+\sigma^{(2)}_{acb}$. Note that $\sigma_{x[yz]}^{(2),BCD}$ actually vanishes when the Fermi surface sum is computed.}
    \label{fig:firstsecond order} 
\end{figure*}

There are three separate contributions to the second order conductivity, a Drude conductivity $\sigma^{(2),\text{D}}_{\alpha \beta \gamma}$, a Berry curvature dipole conductivity $\sigma^{(2),\text{BCD}}_{\alpha \beta \gamma}$ and an intrinsic berry connection polarizability conductivity $\sigma^{(2),\text{BCP}}_{\alpha \beta \gamma}$. From Eq.~\eqref{Tconst} we see that the only surviving contributions are of the form $\sigma_{x[yz]}$ up to permutations, where $[...]$ denotes index symmetrization. This forces the BCD contribution to sum to zero. As in the linear case, all of these quantities can be expressed as Fermi surface integrals (see App.~\ref{app:condFS}). In Fig.~\ref{fig:firstsecond order}, we plot the distribution of the different contributions to the second order conductivity over the Fermi pocket centered at $P$.
\subsubsection{Third order response}
\begin{figure*}[h!] 
    \centering 
    \includegraphics[width=0.49\linewidth]{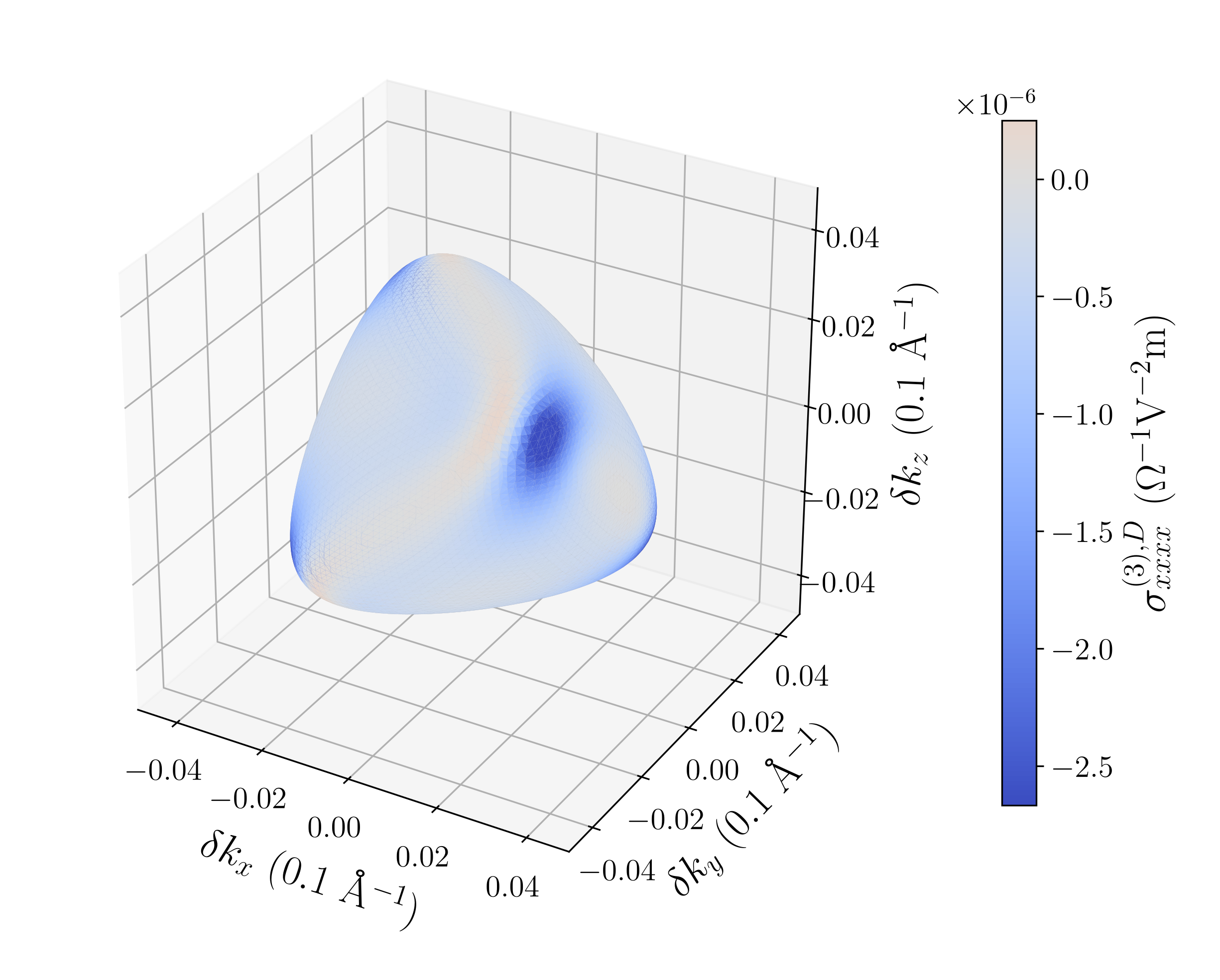} 
    \includegraphics[width=0.49\linewidth]{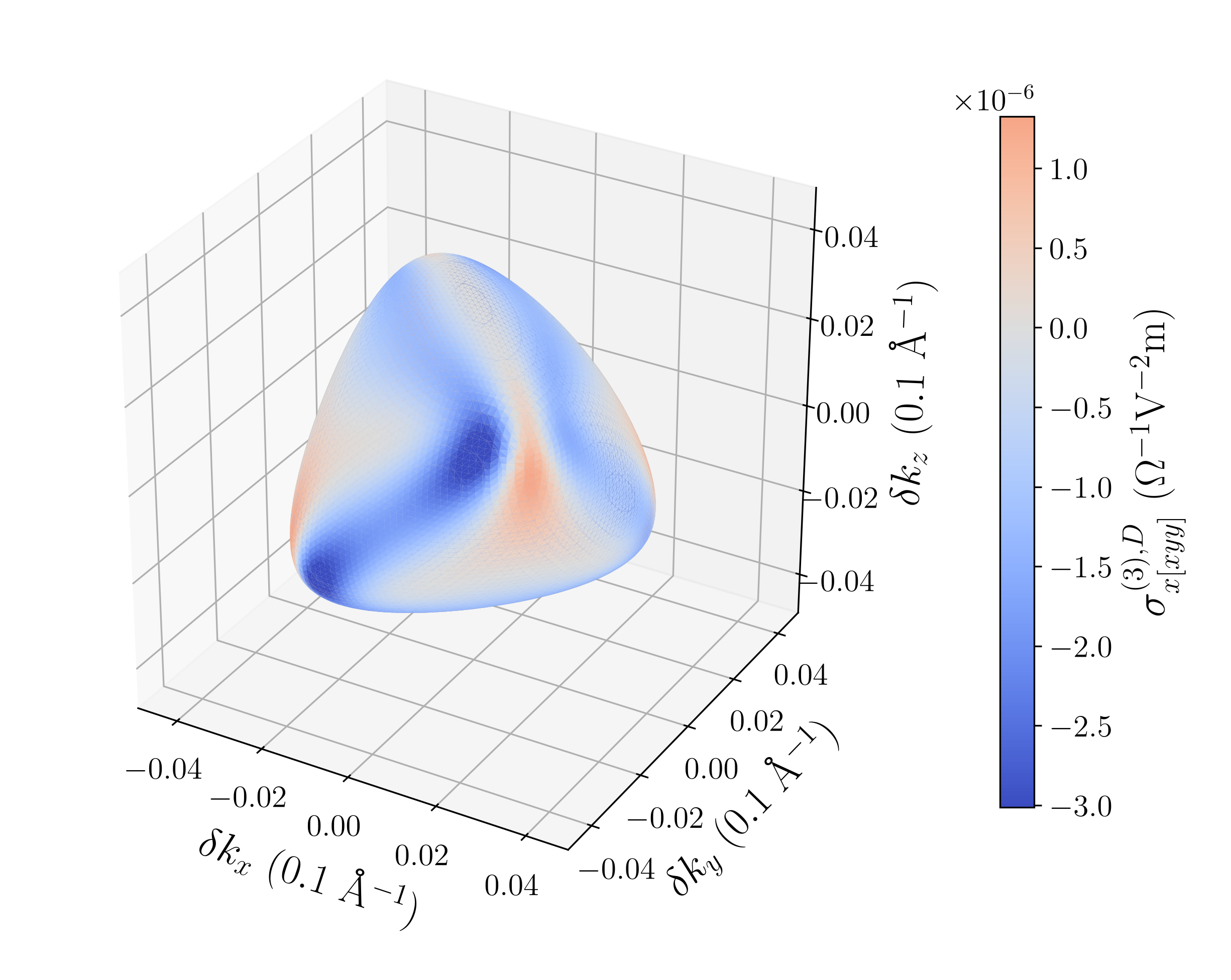}
    \includegraphics[width=0.49\linewidth]
    {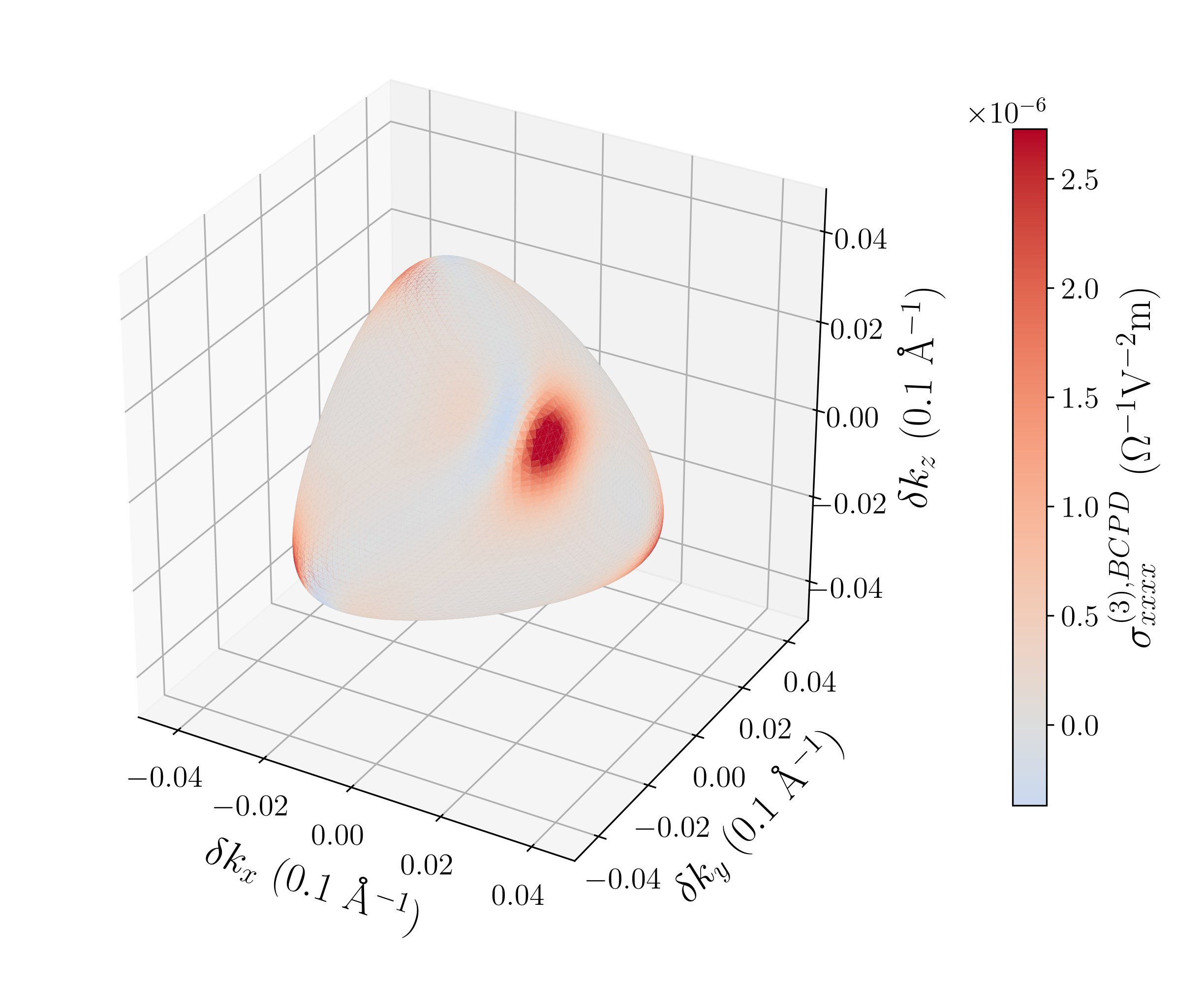} 
    \includegraphics[width=0.49\linewidth]{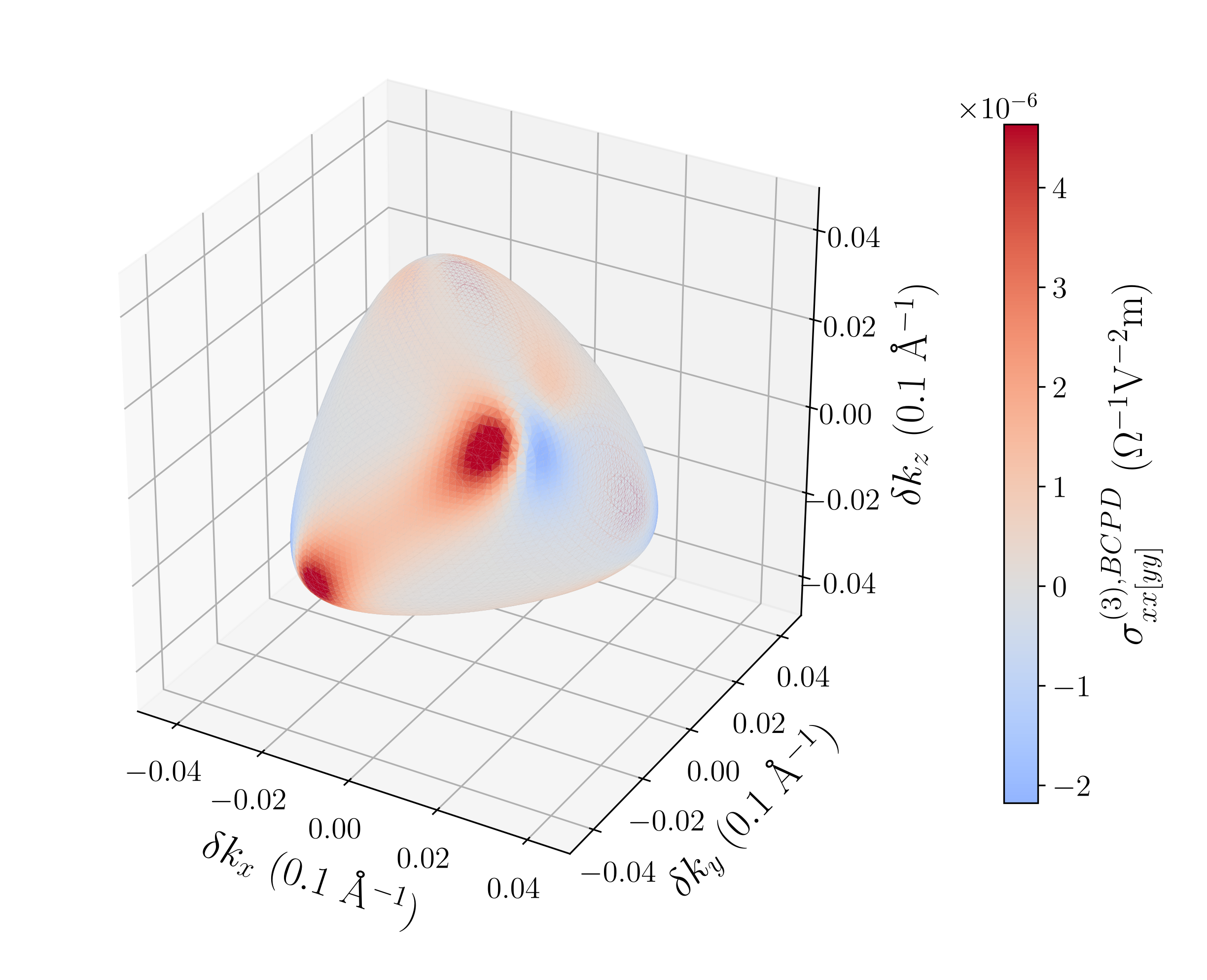}
    \includegraphics[width=0.49\linewidth]{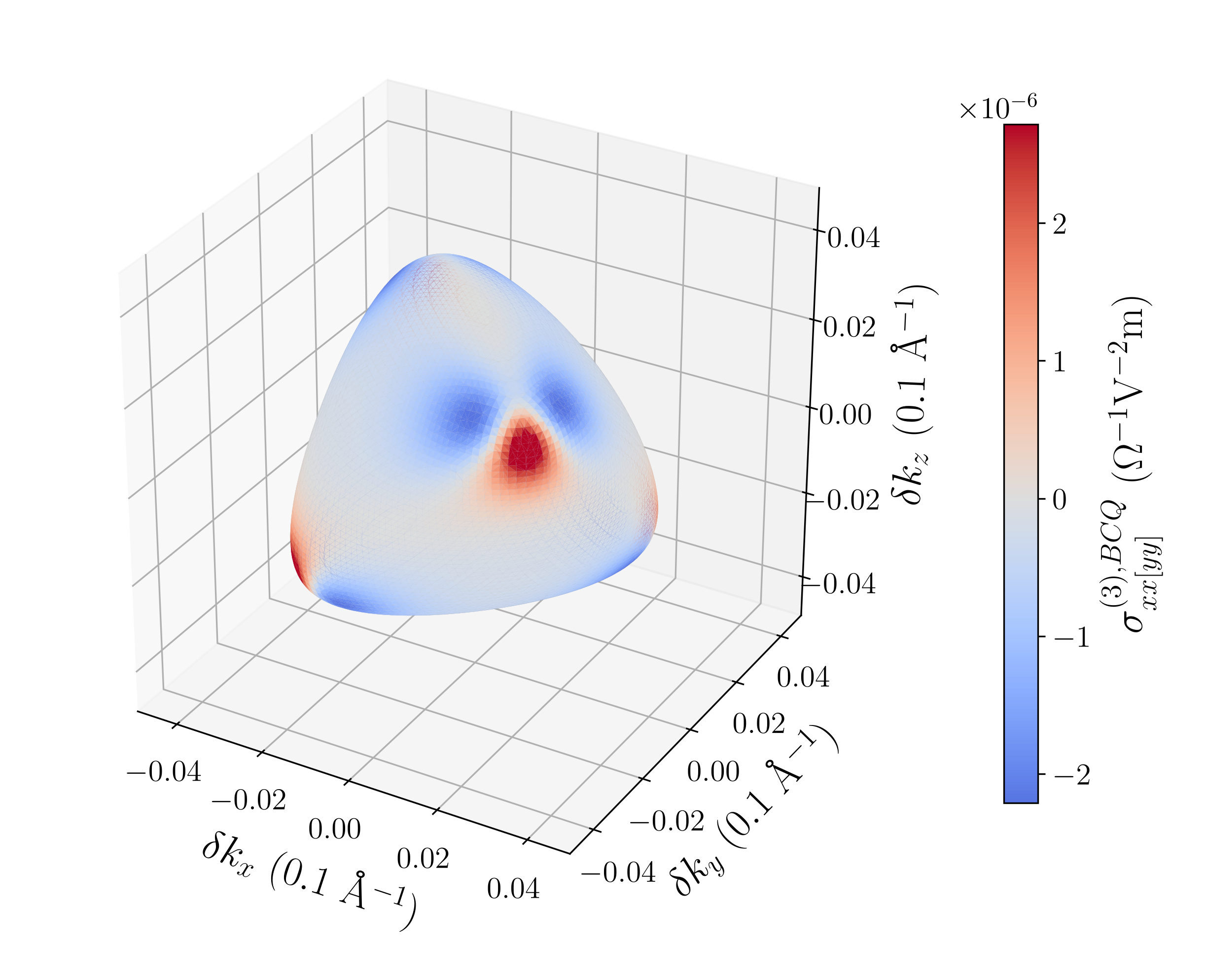}
    \includegraphics[width=0.49\linewidth]{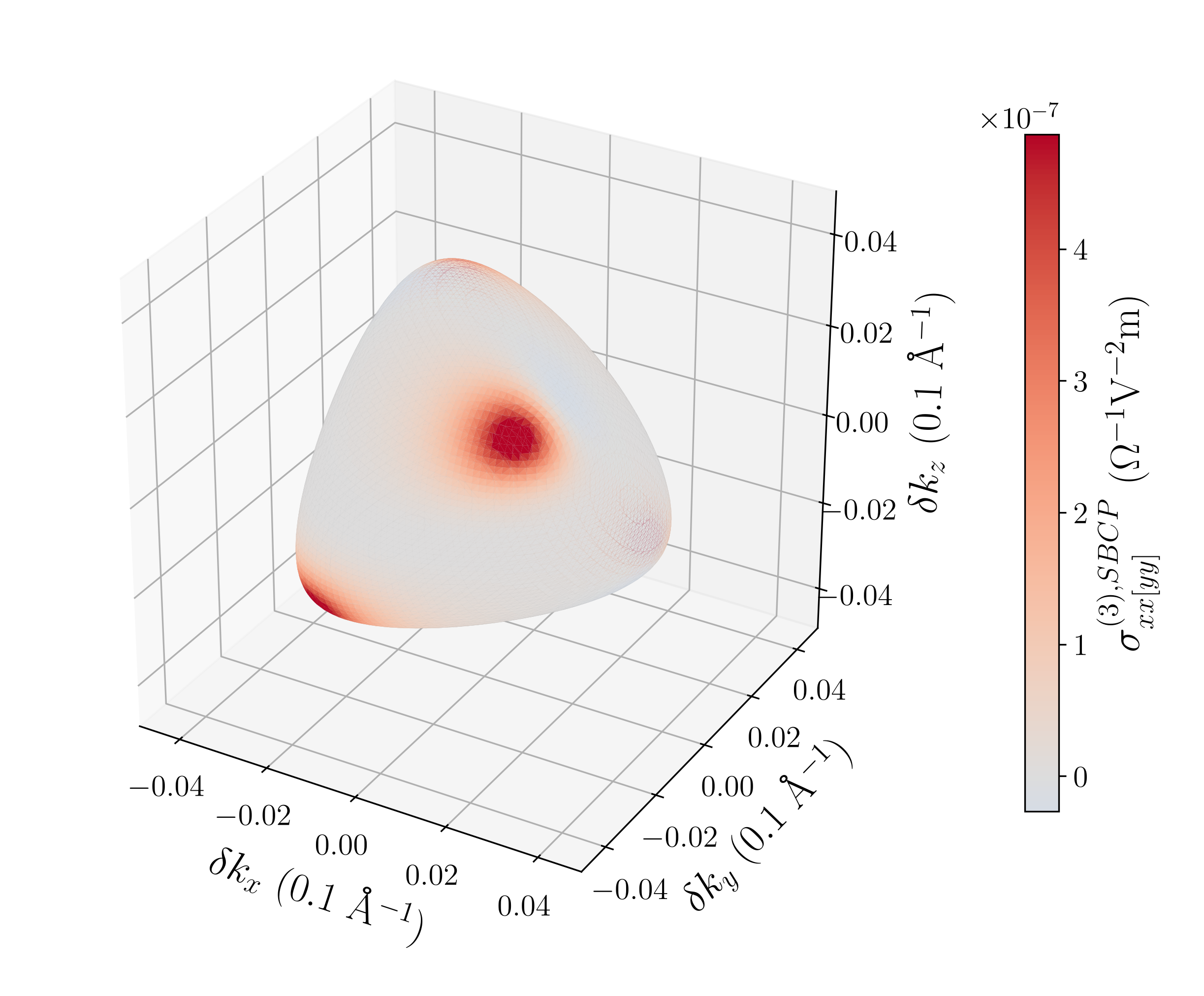}
    \caption{$\textbf{k}$-resolved contributions to the third order conductivity on the Fermi pocket centered at $P$, coming from the third order Drude conductivity $\sigma^{(3),\text{D}}_{\alpha \beta \gamma \eta}$, the Berry curvature quadrupole conductivity $\sigma^{(3),\text{BCQ}}_{\alpha \beta \gamma \eta}$, the Berry connection polarizability dipole $\sigma^{(3),\text{BCPD}}_{\alpha \beta \gamma \eta}$ and the intrinsic second berry connection polarizability conductivity $\sigma^{(3),\text{SBCP}}_{\alpha \beta \gamma \eta}$. When summed over the entire Fermi surface, these give $c_i$ in Eq.~\eqref{jEgeneral1}. The only contributions which can sum to a finite value are of type $\sigma^{(3)}_{aaaa}$ and $\sigma^{(3)}_{a[abb]}$ where $[...]$ denotes index symmetrization (i.e.~$\sigma^{(3)}_{a[abb]} = \sigma^{(3)}_{aabb}+\sigma^{(3)}_{abab}+\sigma^{(3)}_{abba}$). Note that the SBCP and BCQ contribution to $\sigma^{(3)}_{xxxx}$ are pointwise zero on the entire Fermi surface (so they are not plotted here).}
    \label{fig:third order} 
\end{figure*}

There are four separate contributions to the third order conductivity, a Drude conductivity $\sigma^{(3),\text{D}}_{\alpha \beta \gamma \eta}$, a Berry curvature quadrupole conductivity $\sigma^{(3),\text{BCQ}}_{\alpha \beta \gamma \eta}$, a Berry connection polarizability dipole conductivity $\sigma^{(3),\text{BCPD}}_{\alpha \beta \gamma \eta}$ and an intrinsic second berry connection polarizability conductivity $\sigma^{(3),\text{SBCP}}_{\alpha \beta \gamma \eta}$. As in the linear case, all of these quantities can be expressed as Fermi surface integrals (see App.~\ref{app:condFS}). In Fig.~\ref{fig:third order} we plot the distribution of the different contributions to the third order conductivity over the Fermi pocket centered at $P$.

\section{Classification of multifold fermions}\label{app:multifold_tables}
We summarize the classification of multifold fermions, first obtained in Ref.~\onlinecite{bradlyn2016beyond}, in Table~\ref{tab:3fold}, \ref{tab:4fold}, \ref{tab:sixfold}, \ref{tab:eightfold}. For each type of multifold crossing, we list the corresponding space group, crystal point group, $k$ point at which the crossing occurs, and the little co-group at that $k$ point. 
\begin{table*}[h]
  \centering
  \caption{Three-fold fermions (i.e. spin-1) in materials with SOC and time-reversal symmetry. These only occur in non-centrosymmetric, non-symmorphic space groups. Centrosymmetric space groups would lead to a six-fold fermion.}
  \label{tab:3fold}
  \setlength{\tabcolsep}{9pt}
\renewcommand{\arraystretch}{1.1}
  \begin{tabular}{ccccc}
    \hline\hline
   Space group & Crystal point group & $k$-point & TRIM & Little co-group at $k$ \\
    \hline
I$2_13$ (199) & $T$&
      $P$ & \text{\sffamily X} & $T$ \\
I$4_123$ (214) & $O$ &
      $P$ & \text{\sffamily X} & $T$ \\
I$\bar{4}3$d (220) & $T_d$ &
      $P$ & \text{\sffamily X} & $T$ \\
    \hline\hline
  \end{tabular}
\end{table*}

\begin{table*}[h]
  \centering
  \caption{Four-fold fermions (we only consider spin-$\frac{3}{2}$ type) with SOC and time-reversal symmetry. These only occur in non-centrosymmetric, symmorphic and non-symmorphic space groups. Centrosymmetric space groups would either lead to an eight-fold crossing, or Dirac-like four-fold crossings (i.e. two doubly-degenerate bands).}
  \label{tab:4fold}
    \setlength{\tabcolsep}{9pt}
\renewcommand{\arraystretch}{1.1}
  \begin{tabular}{ccccc}
    \hline\hline
    Space groups & Crystal point group & $k$-point & TRIM & Little co-group at $k$ \\
    \hline
    195–199 & $T$ & $\Gamma$ & \checkmark & $T$ \\
    207–214 & $O$  & $\Gamma$ & \checkmark & $O$ \\
    207, 208 & $O$ & $R$ & \checkmark &$O$ \\
    211, 214 & $O$  & $H$ & \checkmark & $O$ \\
    215-220 & $T_d$ & $\Gamma$ & \checkmark & $T_d$\\
    $P\bar{4}3m$ (215) & $T_d$ & $R$ & \checkmark & $T_d$\\
    $I\bar{4}3m$ (217) & $T_d$ & $H$ & \checkmark & $T_d$\\
    $I\bar{4}3m$ (217) & $T_d$ & $P$ & \text{\sffamily X} & $T_d$ \\
    \hline\hline
  \end{tabular}
\end{table*}

\begin{table*}[h]
  \centering
  \caption{Six-fold fermions (Kramers pair of 3-fold nodes due to time-reversal at TRIMs or inversion plus time-reversal at non-TRIMs) with SOC and time-reversal symmetry. These occur in non-symmorphic, centrosymmetric and non-centrosymmetric space groups only.}
  \label{tab:sixfold}
    \setlength{\tabcolsep}{9pt}
\renewcommand{\arraystretch}{1.1}
  \begin{tabular}{ccccc}
    \hline\hline
   Space group & Crystal point group & $k$-point & TRIM & Little co-group at $k$ \\
    \hline
 P$2_13$ (198) & $T$ &
      $R$ & \checkmark & $T$ \\
P$4_332$ (212) & $O$ &
      $R$ & \checkmark & $O$ \\
P$4_132$ (213) & $O$ &
      $R$ & \checkmark & $O$ \\
Pa$\bar{3}$ (205) & $T_h$ &
      $R$ & \checkmark & $T_h$ \\
    Ia$\bar{3}$ (206) & $T_h$ &
      $P$ & \text{\sffamily X} & $T_h$ \\
 Ia$\bar{3}d$ (230) & $O_h$  &
      $P$ & \text{\sffamily X} & $O_h$ \\
    \hline\hline
  \end{tabular}
\end{table*}

\begin{table*}[h]
  \caption{Eight-fold fermions (Kramers pair of 4-fold nodes)  with SOC and time-reversal symmetry. These occur in non-symmorphic, centrosymmetric and non-centrosymmetric space groups only.}
  \label{tab:eightfold}
    \setlength{\tabcolsep}{9pt}
\renewcommand{\arraystretch}{1.1}
  \begin{tabular}{ccccc}
    \hline\hline
    Space group & Crystal point group
    & $k$-point & TRIM & Little co-group at $k$ \\
    \hline
    P$4/ncc$ (130)  & $D_{4h}$ &
      $A$ & \checkmark & $D_{4h}$ \\
    P$4_2/mbc$ (135) & $D_{4h}$&
      $A$ & \checkmark & $D_{4h}$ \\
    Pn$\bar{3}$n (222) & $O_h$  &
      $R$ & \checkmark & $O_h$ \\
    Pm$\bar{3}$n (223)& $O_h$  &
      $R$ & \checkmark & $O_h$ \\
    Ia$\bar{3}d$ (230) & $O_h$  &
      $H$ & \checkmark & $O_h$ \\
    P$\bar{4}3$n (218) & $T_d$ &
      $R$ & \checkmark & $T_d$ \\
    I$\bar{4}3$d (220) & $T_d$  &
      $H$ & \checkmark & $T_d$ \\
    \hline\hline
  \end{tabular}
\end{table*}
\end{appendix}
\end{document}